\documentclass[superscriptaddress,prd,aps,amssymb,nofootinbib,twocolumn,showkeys,showpacs]{revtex4}

\usepackage{graphicx, wrapfig, epsfig}

\begin{document}
\preprint{}

\title{Study of the Spectrum of Inflaton Perturbations}

\author{Matthew M. Glenz}
\email{mmglenz@uwm.edu}
\affiliation{ Department of Physics,
University of Wisconsin-Milwaukee, P.O. Box 413, Milwaukee, WI
53201}

\author{Leonard Parker}
\email{leonard@uwm.edu}
\affiliation{ Department of Physics,
University of Wisconsin-Milwaukee, P.O. Box 413, Milwaukee, WI
53201}
\date{May 15, 2009}

\begin{abstract}

We examine the spectrum of inflaton fluctuations
resulting from any given long period of exponential inflation.  
Infrared and ultraviolet divergences in the inflaton
dispersion summed over all modes do not appear in our
approach. We show how the scale-invariance of the perturbation spectrum arises.  We also examine the spectrum of scalar
perturbations of the metric that are created by the inflaton
fluctuations that have left the Hubble sphere during inflation
and the spectrum of density perturbations that they produce
at reentry after inflation has ended.  When the inflaton dispersion
spectrum is renormalized during the expansion, we show
(for the case of the quadratic inflaton potential) that the density perturbation spectrum approaches a mass-independent limit as 
the inflaton mass approaches zero, and remains near 
that limiting value for masses less than about $1/4$ of 
the inflationary Hubble constant. We show that this limiting behavior does not occur if one only makes the Minkowski space subtraction,
without the further adiabatic subtractions that involve time derivatives
of the expansion scale factor $a(t)$. We also find a parametrized
expression for the energy density produced by the change
in $a(t)$ as inflation ends. If the end of inflation were sufficiently
abrupt, then the temperature corresponding to this energy
density could be very significant.  We also show that fluctuations
of the inflaton field that are present before inflation starts are not
dissipated during inflation, and could have a significant observational
effect today. The mechanism for this is caused by the initial 
fluctuations through stimulated emission from the vacuum.

\end{abstract}

\maketitle

\section{Introduction}
\label{sec:intro}

Cosmological inflation predicts an amplification of quantum
fluctuations that perturbs the background homogeneity of the
universe and leads to regions of spacetime of nonuniform density
\cite{SR}. The dispersion of the quantized inflaton fluctuation
field, $\delta\phi$, is taken as a measure of the inhomogeneity of
the field during inflation. This inhomogeneity perturbs the
gravitational field during inflation, and its perturbations set the
initial conditions for the acoustic oscillations of the plasma and
matter that are present after the inflationary era.

The dispersion of $\delta\phi$ has an ultraviolet divergence
that results from integrating over all modes at a given time
during inflation. As the highest frequency modes correspond
to wavelengths that never exit the de Sitter horizon $H^{-1}$ during
inflation, one may think that such modes can be ignored or dealt with by
means of a cut-off or by standard curved-spacetime regularization
and renormalization methods without observable consequences.
However, \cite{Parker1} showed
that when the method of adiabatic regularization is used to
renormalize the dispersion of  $\delta\phi$, the dispersion
spectrum at wavelengths that have left the de Sitter horizon is
{\it significantly} affected.
Adiabatic regularization in Robertson-Walker universes has
been shown \cite{BiLuPiJuAn} to give the same result as other
forms of regularization and renormalization, including point-splitting
regularization; and the adiabatic condition has been shown to be
closely related to the Hadamard condition in curved spacetime.
Because of the time-translation properties of the universe during inflation,
renormalization of the spectrum of inflaton fluctuations does not alter
the near scale-invariance of the spectrum of perturbations
as they enter the Hubble horizon of the post-inflationary universe.
However,  renormalization may alter the relation between the magnitude
of the inflaton perturbation spectrum and the implied value
of $H$ during inflation. This would be of importance in testing
theories that predict the value of $H$ directly.  The effects studied
in \cite{Parker1} have been further elucidated in
\cite{Agullo, Agullo2}.

Is there a way to obtain the dispersion spectrum of inflaton
fluctuations
by using {\em only} the well-tested techniques of quantum field
theory in
Minkowski space without appealing to renormalization in curved spacetime?
Such a method was developed by one of the authors (LP) in the early
1960's in the first part of his Ph.~D. thesis \cite{Parker2}.  
First we explain how and why the method works. 
Then we trace the evolution of the quantized inflaton fluctuation 
field on the background metric, without considering the
metric perturbations and the density perturbations that they
lead to after reheating.  In that case, we can use the method
to obtain the dispersion spectrum and to
examine the scale invariance of the spectrum resulting from a long
period of inflation. We also demonstrate how the method avoids
{\it both} infrared and ultraviolet divergences in the dispersion of the
inflaton flucutation field. 

Then we turn to processes involving the production of
scalar perturbations of the metric by the modes of the inflaton
fluctuation field after they exit the inflationary Hubble sphere; and
to the spectrum of initial density perturbation they create upon
reentry into the Hubble sphere after the end of inflation.  For
these considerations we consider the consequences of using
only the Minkowski space regularization, as compared with
regularizing infinities of the inflaton dispersion that involve 
derivatives of $a(t)$ duriing the expansion, and using the
regularized inflaton dispersion at the time when they induce
significant metric perturbation after leaving the inflating
Hubble horizon.  We find that there are significant differences
in the magnitudes of the density perturbations and their
dependence on the effective mass of the inflaton fluctuation field.
In particular, using the regularized dispersion leads 
to initial density perturbations that approach a nonzero value
in the limit of zero inflaton mass, and that are almost independent of the inflaton mass if that mass is less than a significant fraction
of the inflationary Hubble constant.

In \cite{Parker2}, LP showed that an expanding universe creates
elementary particles. He assumed that the quantized field 
is evolved by
the generally covariant field equation in the expanding universe.
If the quantized particle field was  expanded in
terms of mode function solutions of the field equation and the creation
and annihilation operators were defined as usual in terms of the
coefficients of the mode functions, then he found that
the particle number density created in a given mode was finite, but when
summed over all modes the particle number density had an ultraviolet
divergence.  This raised the questions, (1) should the particle number operator
that yields the particle number per unit physical volume be renormalized
during the expansion of the universe, and (2) is it possible to prove,
using only the known and tested quantum field theory in Minkowski space,
that the density of particles created during the expansion of the universe,
as predicted by quantum field theory, is actually finite?

To answer the latter question, he
considered a general expansion\footnote{The word
``expansion'' is used in a general sense, to include the possibility
of contraction.} of the universe that started smoothly from Minkowski
space in the distant past and after expanding
in an arbitrary smooth manner approached a Minkowski space again
in the distant future. It is reasonable to assume that the density of
real particles created during the expansion is not greatly disturbed
by the gradual and smooth joining to the early- and late-time
Minkowski spaces. He showed that for such asymptotically Minkowskian
expansions of the universe, the number density of particles created by the
expansion of the universe was finite. {\it No renormalization of the particle
number in the curved spacetime during the expansion of the universe
was used to obtain this result.} Only the standard definition of
particle number in the initial and final Minkowski spaces
was used.  He employed general mathematical theorems to
prove that the particle number in a given co-expanding volume is
an adiabatic invariant, and that the total number of particles created
in the co-expanding volume by the expansion of the universe from
the initial to the final Minkowski space is finite when summed
over all modes.  This was proved by considering the Bogoliubov
transformation,
$a_{\vec{k}}=\alpha_{k}A_{\vec{k}}+\beta_{k}^{*}A_{-\vec{k}}^{\dagger}$,
that relates the annihilation operator $a_{\vec{k}}$ for particles in mode
$\vec{k}$ at early times to the annihilation and creation operators at
late times. He showed that the quantity $|\beta_k|^2$ that determines
the average number of particles, created in mode $\vec{k}$ from the
early-time vacuum by the expansion of the universe, vanishes
faster than any inverse power of $k$, as $k \rightarrow \infty$.
Consequently, the average number of created particles,
summed over all modes, is finite as measured in the late-time
Minkowski space.  This result is independent of renormalization
in curved spacetime.

He addressed the former question (about renormalization of
the particle number operator during the expansion) by considering the
properties of a device that measures the number of particles per
unit physical volume.  As shown in detail in \cite{Parker2} and
summarized in \cite{Parker3}, the natural assumption that the
measuring instrument is not able to measure a particle number that
has very fast small oscillations, together with the requirement that
the number measured in a given physical volume should be an integer,
leads to a renormalized definition of the Hermitian number operator
corresponding to the quantity actually measured by such an instrument.
This physically relevant renormalized number operator in an expanding
universe is the ``bare'' number operator obtained from the mode function
expansion of the field, renormalized by making the adiabatic subtractions
to second adiabatic order. The method of adiabatic regularization was
further developed and applied to the energy-momentum tensor
with Fulling and Hu in \cite{Parker5}.

In order to obtain unambiguous results, independent of
regularization and renormalization in curved spacetime, for the
dispersion of the inflaton field resulting from a long period of
inflation, we will consider a model of the inflationary universe
that starts smoothly from Minkowski space in the distant past and,
after undergoing any given number of e-foldings of inflation,
approaches a Minkowski space again in the distant future.
(We do not take up perturbations of the background metric 
that are produced by the created inflaton fluctuations 
until later in this paper.) We first take
the initial state of the quantized inflaton perturbation field to be
the Minkowski vacuum. 
The asymptotic Minkowski space
in the distant past may be regarded as a convenient way of
specifying unambiguous initial conditions that can be interpreted
without reference to curved spacetime. The asymptotic Minkowski
space at late times plays a similar role in permitting us to
unambiguously interpret the spectrum of inflaton perturbations at
late times after any given number of e-foldings of inflation. As
explained above in our discussion of \cite{Parker2, Parker3}, we
must do the joining at early and late times as smoothly as possible
to obtain results that do not have an ultraviolet divergence when
summed over all modes.
Our present method, in which $a(t)$ is asymptotically flat at
early times, also avoids infrared divergences, such as those
that were found in the treatment of
graviton production from inflation by \cite{Allen}.  The
infrared and ultraviolet divergences found by \cite{Yajnik}
are also absent.

In Section~\ref{sec:compos}, we specify the inflationary spacetime
by joining together a composite scale factor built of segments for
which the solution to the evolution equation of the inflaton
fluctuation field is known analytically.  Because observation
\cite{3WMAP, 5WMAP} favors a period of nearly exponential inflation
with a slowly changing inflationary Hubble constant, $H_{\rm infl}$,
we include as the middle part of our composite scale factor a region
of exponential inflation.  We take $H_{\rm infl}$ as a constant, but
we can incorporate a slow change by using the usual adiabatically
adjusted solution in which the mode-function solutions are adjusted
by neglecting the time-derivatives, but including the slow change of
$H_{\rm infl}$. The initial and final asymptotically Minkowskian
segments have a number of adjustable parameters that allow us to
choose any value of $H_{\rm infl}$ and any number of e-folds of
inflation, while joining the scale factor $a(t)$ continuously and
with continuous first and second derivatives. That is, $a(t)$ will
be a $C^2$ function. This is the minimum degree of continuity of
$a(t)$ that, in general, will give a finite energy density in the
late-time Minkowski space. We can also adjust parameters that
determine how quickly inflation ends. 
In Section~\ref{sec:evo}, we obtain the asymptotic conditions on the
modes of the inflaton perturbation field. 

In Section~\ref{sec:number}, we discuss the average 
number of quanta of
the inflaton perturbation field for a pure state (the early-time
vacuum) and for a statistical mixture of states having different
numbers of particles present at early times. We discuss a surprising
feature that results from the stimulated creation of particles if
there are particles present at early times. This effect remains
significant at late times, particularly at scales that may be
relevant to the large scale structure of the universe.
Without amplification by stimulated emission, the initial inflaton
fluctuations (i.e., particles or quanta of the inflaton fluctuation
field) would be dispersed by the inflationary expansion and
would have negligible effect. We will
take this up further in a later paper.
In standard treatments of inflation,
it is assumed that the initial conditions have no significant
effect after a sufficient number of e-foldings.  This statement,
although it seems intuitively obvious, is not correct unless the
initial state of the inflaton field is the vacuum state. 

In Section~\ref{sec:match}, we solve the evolution equation for
quantized fluctuations of the inflaton field for the above class of
scale factors $a(t)$. We do this by matching the analytic solutions
for the modes of the inflaton field perturbations, $\delta\phi$, and
their time-derivatives at the joining points where we spliced
together the different segments of the scale factor.  We introduce
the general evolution equation for $\delta\phi$ with a constant
effective mass and focus initially on the exact solutions of the
massless case.  In Section~\ref{sec:masslesspc}, we find the number
of particles created in each mode for inflaton fluctuations of $0$
effective mass, and we discuss the effect of discontinuities in a(t)
and its derivatives on the average number of inflaton perturbation
quanta created in each mode.

In Section~\ref{sec:MasslessDispersionSpectrum}, we obtain the
dispersion spectrum of the massless inflaton fluctuation field.
As mentioned
above, there are no infrared or ultraviolet divergences in these
quantities. We also find effects that are evident in our
dispersion spectrum and depend on the phases of the inflaton
perturbation modes at the time that inflation begins in our model.

In Sections~\ref{sec:massivepc} and~\ref{sec:mass}, we find the
corresponding results for quantized inflaton fluctuations of
non-zero effective mass. In Section~\ref{sec:spec}, we discuss the
dependence of the spectral index on the effective mass of
$\delta\phi$.  We also investigate the effect of the total duration
of inflation on the scale-invatiance of very long wavelength
perturbation modes.  In Section~\ref{sec:dens}, we discuss the
initial density perturbations we would expect to be produced
after reheating, and we show that 
there is a significant difference when we employ
renormalization of the inflaton dispersion
during the expansion, as compared
with employing only the Minkowski space renormalization.
In Section \ref{sec:reheat}, we use our
parametrized scale factor to calculate the contribution of the
changing gravitational field to reheating (see also \cite{Ford}).

\section{Composite Scale Factor}
\label{sec:compos}

We consider the metric
\begin{equation}
ds^{2}=dt^{2}-a^{2}(t)((dx)^{2}+(dy)^{2}+(dz)^{2}).
\label{eq:metric}
\end{equation}
The time $t$ will run continuously from $-\infty$ to $\infty$. The
scale factor $a(t)$ will be composed of three segments.  Our
scale factor will generally be $C^2$, i.e., a continuous
function with continuous first and second derivatives everywhere,
including at the joining points between segments.\footnote{We also
briefly consider scale factors that are only $C^1$ or $C^0$ at the
joining points.} The initial and final segments are asymptotically
Minkowskian in the distant past and future, respectively. The middle
segment is an exponential expansion. We choose specific forms for
$a(t)$ in these segments that have exact solutions of the evolution
equations for inflaton quantum fluctuations of zero effective mass.

We emphasize that the initial and final asymptotically flat
regions permit us
to unambiguously interpret our results for free fields without
having to perform any renormalization in curved spacetime.
The final asymptotically flat region will not significantly affect
the result obtained for the spectrum of inflaton perturbations
created by the inflationary segment of the expansion. The
initial asymptotically flat region should have a negligible effect
on the spectrum resulting from a long period of
inflation.\footnote{We do find remnants of the early initial
conditions in the late-time inflaton dispersion spectrum, which
we discuss later.}  If there are cases in which no inflaton
perturbations are created by the period of exponential
inflation, then the inflaton perturbations would
result from the initial asymptotically flat region, possibly
amplified by the long period of inflation .

\begin{figure}[hbtp]
\includegraphics[scale=1.4]{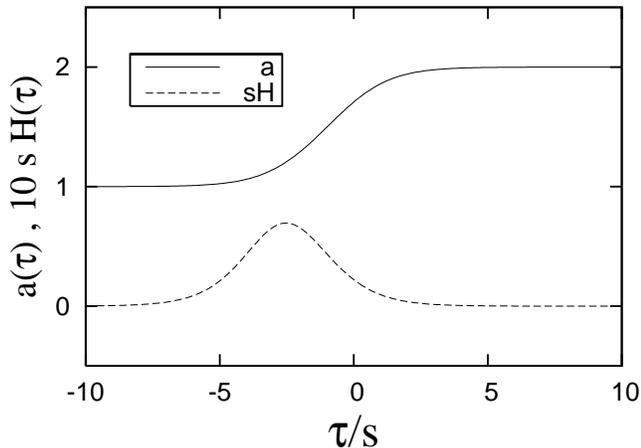}
\caption{\label{fig:aAndHvTau} Scale factor,
$a\left(t(\tau)\right)$, and dimensionless Hubble parameter, $s
H(t(\tau))= s a^{-1}{da/dt}=sa^{-4}{da/d\tau}$, of
Eq.~(\ref{eq:aHG}) with $a_{1}=1$, $a_{2}=2$, $b=0$, and $s=1$. Note
in the graph that the maximum of $H$ occurs at a value of
$a(t(\tau))$ closer to $a_{1}$ than to $a_{2}$.  In both the case
where $a_{2}\gg a_{1}$ and the case where $a_{2}\simeq a_{1}$,
$H_{\rm max}$ occurs at a value of the scale factor where
$a(t(\tau))\simeq a_{1}$. The relationship between proper time and
$\tau$-time is given by Eq.~(\ref{eq:tvtau}).}
\end{figure}

We base each asymptotic segment on a scale factor of the form,
\begin{eqnarray}
&&\!\!\!\!\!\!\!\!\!\!\!\!\!a(t(\tau))=
\nonumber\\
&&\!\!\!\!\!\!\!\!\!\!\!\!\!\bigg\{a_{1}^{\ 4}+e^{\tau/s}[(a_{2}^{\ 4}-a_{1}^{\
4})(e^{\tau/s}+1)+b](e^{\tau/s}+1)^{-2}\bigg\}^{\frac{1}{4}} ,
\label{eq:aHG}
\end{eqnarray}
where $\tau$ is related to the proper time $t$, as defined through
Eq.~(\ref{eq:metric}), by
\begin{equation}
d\tau\equiv a(t)^{-3}dt.
\label{eq:tvtau}
\end{equation}

\begin{figure}[hbtp]
\includegraphics[scale=1.4]{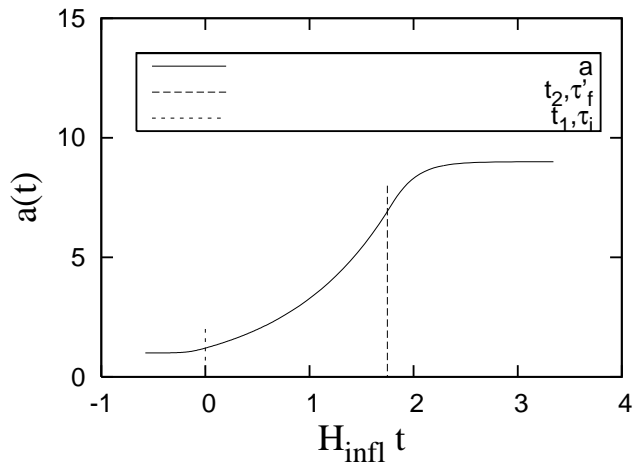}
\caption{\label{fig:diagram} Scale factor plotted versus
dimensionless time.  This illustrative example summarizes our
notation using a moderate expansion of $\sim2$ e-folds. The scale
factor, $a(t)$, is continuous, as are $\dot{a}(t)$ and
$\ddot{a}(t)$.  In this case, the parameters for the initial
asymptotically flat segment are $a_{1i}=1$, $a_{2i}=2$, and
$s_{i}=1$.  The free parameters of the final asymptotically flat
segment are $a_{2f}=9$ and $a_{1f}=6$. The asymptotically flat scale
factor of the initial region joins the exponentially expanding scale
factor of the middle region at a time $t_{1}$ in $t$-time and
$\tau_{i}$ in $\tau$-time. The exponentially expanding scale factor
of the middle region joins the asymptotically flat scale factor of
the final region at a time $t_{2}$ in $t$-time and $\tau'_{f}$ in
$\tau'$-time of the final segment, where a prime is used to
distinguish between the $\tau$-times of the initial and final
segments.}
\end{figure}

The form of the scale factor in Eq.~(\ref{eq:aHG}) is based on the
form of the index of refraction used by Epstein to model the
scattering of radio waves in the upper atmosphere and by Eckart to
model the potential energy in one-dimensional scattering in quantum
mechanics \cite{EE}. It was first used in the cosmological context
by Parker  \cite{CinNature} to model $a(t)$. As can be seen from
Fig.~\ref{fig:aAndHvTau}, this scale factor approaches the constant
$a_{1}$ at early times and the constant $a_{2}$ at late times, and
the constant $s$ determines roughly the interval of $\tau$-time for
$a(t)$ to go from $a_1$ to $a_2$. (A sufficiently large magnitude of
$b$ would produce a bump or valley in $a(t)$.) The parameters
$a_{1}$, $a_{2}$, $b$, and $s$ are different in the initial and
final asymptotically flat segments.  Where confusion would arise we
will include subscripts $i$ in the initial set of parameters and $f$
in the final set of parameters. The equation for the middle
(inflationary) segment of our composite scale factor is given in
terms of proper time by
\begin{equation}
a(t)=a(t_1) e^{H_{\rm infl}(t-t_1)}, \label{eq:infl}
\end{equation}
where $H_{\rm infl}$ is the constant value of $H(t)\equiv a^{-1}
da/dt$ during the exponential expansion of the middle segment.

We define the quantity $N_{e}\equiv \ln\left(a_{2f}/a_{1i}\right)$.
When there is a long period of exponential growth, $N_{e}$ is
essentially the number of e-foldings of inflation. Typically,
$N_{e}$ will be about $60$. Within the final asymptotically flat
scale factor, the ratio of $a_{2f}$ to $a_{1f}$ determines how
gradually the exponential expansion transitions to the
asymptotically flat late-time region. (For example, this ratio might
be 1 e-fold, which we would consider to be relatively gradual, or it
might be 1.0001, which we would consider to be relatively abrupt.)

With our choices of $a(t)$ in the three segments, we
are able to join them so that $a(t)$ and
its first and {\em second} derivatives with respect to time
are continuous everywhere.  This requires that we join the
exponentially expanding segment, in which $H(t)$ has the
constant value $H_{\rm infl}$, to the initial and final segments
at the times when $H(t)$ is a maximum. This maximum value
must equal $H_{\rm infl}$.    A simple power law form of the
scale factor, such as that of a radiation-dominated universe, could
not be used to simultaneously maintain the continuity of the scale
factor and its first and second derivatives when matched directly to
the inflationary segment of exponential expansion.

With $b_{i}=0$ and $b_{f}=0$, we then find the following
expressions. The time $\tau_{i}$ at which the first segment joins to
the exponential segment is
\begin{equation}
\tau_{i}=s_{i}\ln\left(\frac{3a_{1i}^{\ \ 4}-3a_{2i}^{\ \
4}+C_{i}}{8a_{2i}^{\ \ 4}}\right).
\end{equation}
The constant $a(t_{1})$ in Eq.~(\ref{eq:infl}) is
\begin{equation}
a(t_{1})=\left(\frac{-3a_{1i}^{\ \ 4}-3a_{2i}^{\ \
4}+C_{i}}{2}\right)^{1/4}.
\end{equation}
Because the maximum value of $H(t)$ in the first segment
must equal $H_{\rm infl}$, we find that
\begin{eqnarray}
H_{\rm infl}=&&\left[\frac{2^{3/4}\left(-a_{1i}^{\ \ 4}+a_{2i}^{\ \
4}\right)}{a_{2i}^{\ \ 4}\left(11a_{1i}^{\ \ 4}-3a_{2i}^{\ \
4}+C_{i}\right)^{2}s_{i}}\right]
\nonumber\\
&&\times\left(-3a_{1i}^{\ \ 4}-3a_{2i}^{\ \ 4}+C_{i}\right)^{1/4}
\nonumber\\
&&\times\left(3a_{1i}^{\ \ 4}-3a_{2i}^{\ \ 4}+C_{i}\right),
\label{eq:Hmax}
\end{eqnarray}
where
\begin{equation}
C_{i}\equiv\sqrt{9a_{1i}^{\ \ 8}+46a_{1i}^{\ \ 4}a_{2i}^{\ \
4}+9a_{2i}^{\ \ 8}}.
\end{equation}

Once we choose values for $a_{1f}$ and $a_{2f}$, the
remaining constants are determined to have the following values:
\begin{eqnarray}
s_{f}=&&\left[\frac{2^{3/4}\left(-a_{1f}^{\ \ 4}+a_{2f}^{\ \
4}\right)}{a_{2f}^{\ \ 4}\left(11a_{1f}^{\ \ 4}-3a_{2f}^{\ \
4}+C_{f}\right)^{2}H_{\rm infl}}\right]
\nonumber\\
&&\times\left(-3a_{1f}^{\ \ 4}-3a_{2f}^{\ \ 4}+C_{f}\right)^{1/4}
\nonumber\\
&&\times\left(3a_{1f}^{\ \ 4}-3a_{2f}^{\ \ 4}+C_{f}\right).
\end{eqnarray}
Denote the parameter $\tau$ of Eq.~(\ref{eq:aHG})
as $\tau'$ in the final segment.  At the time $\tau'_{f}$
when the exponential segment joins to the final segment,
we find that
\begin{equation}
\tau'_{f}=s_{f}\ln\left(\frac{3a_{1f}^{\ \ 4}-3a_{2f}^{\ \
4}+C_{f}}{8a_{2f}^{\ \ 4}}\right).
\end{equation}
The corresponding proper time $t$ at which the
exponential segment joins to the final segment is
\begin{equation}
t_{2}=\frac{1}{4H_{\rm infl}}\ln\left(\frac{-3a_{1f}^{\ \
4}-3a_{2f}^{\ \ 4}+C_{f}}{-3a_{1i}^{\ \ 4}-3a_{2i}^{\ \
4}+C_{i}}\right)+t_{1},
\end{equation}
where
\begin{equation}
C_{f}\equiv\sqrt{9a_{1f}^{\ \ 8}+46a_{1f}^{\ \ 4}a_{2f}^{\ \
4}+9a_{2f}^{\ \ 8}}.
\end{equation}

See Fig.~\ref{fig:diagram} for a schematic diagram of how we match
our segments of the scale factor together.

\section{Early and Late Asymptotic Conditions on $\delta \phi$}
\label{sec:evo}

Consider an inflaton field composed of a spatially homogeneous term
plus a first order perturbation,
\begin{equation}
\phi(\vec{x},t)=\phi^{(0)}(t)+\delta\phi(\vec{x},t).
\end{equation}
We investigate, in units of $\hbar=c=1$, a minimally-coupled scalar
field that obeys the evolution equation:
\begin{eqnarray}
\partial_{t}^{\
2}\delta\phi+3H\partial_{t}\delta\phi-a^{-2}(t)\sum_{i=1}^{3}\partial_{i}^{\
2}\delta\phi+m(\phi^{(0)})^{2}\delta\phi=0.
\nonumber\\
\
\end{eqnarray}
The mass term is related to the inflationary potential by
\begin{equation}
m(\phi^{(0)})^{2}=\frac{d^{2}V}{d(\phi^{(0)})^{2}}.
\label{eq:massterm}
\end{equation}
For simplicity, we take $m(\phi^{(0)})^2$ as a constant, $m^2$.

The quantized field $\delta\phi$ can be written
in terms of the early time creation and annihilation operators,
$A_{\vec{k}}^{\dagger}$ and $A_{\vec{k}}$, as
\begin{equation}
\delta\phi=\sum_{\vec{k}}\left(A_{\vec{k}}f_{\vec{k}}+
A_{\vec{k}}^{\dagger}f_{\vec{k}}^{*}\right)
\equiv \sum_{\vec{k}} \delta {\hat \phi}_{\vec k} ,
\label{eq:phi}
\end{equation}
where\footnote{For simplicity, we are imposing
periodic boundary conditions upon a cubic coordinate volume,
$V=L^{3}$. In the continuum limit $L$ would go to infinity.}
\begin{equation}
f_{\vec{k}}=V^{-\frac{1}{2}}e^{i\vec{k}\cdot\vec{x}}\psi_{k}(t(\tau)).
\label{eq:wave}
\end{equation}
The function $\psi_{k}(t)$ satisfies
\begin{equation}\partial_{t}^{2}\psi_{k}(t)+3H\partial_{t}\psi_{k}(t)+
\frac{k^{2}}{a^{2}(t)}\psi_{k}(t)+m^{2}\psi_{k}(t)=0,
\label{eq:evot}
\end{equation}
where $k=2\pi n/L$, with $n$ an integer. Because
the creation and annihilation operators in Eq.~(\ref{eq:phi})
correspond to particles at early times, we require that
$\psi_{k}$ satisfies the early-time positive
frequency condition
\begin{equation}
\lim_{\tau\rightarrow -\infty}\psi_{k}(t(\tau)) \sim
\frac{1}{\sqrt{2a_{1i}^{3} \omega_{1i}(k)} }
e^{-ia_{1i}^{3}\omega_{1i}(k)\tau},
\label{eq:minkowski}
\end{equation}
where  $\omega_{1i}(k) \equiv \sqrt{(k/a_{1i})^2 + m^2}$.

At late times, this solution will have the asymptotic form
\begin{eqnarray}
\lim_{\tau'\rightarrow \infty}\psi_{k}(t(\tau')) &\sim&
\frac{1}{\sqrt{2a_{2f}^{3} \omega_{2f}(k)} }
\bigg[\alpha_{k}e^{-ia_{2f}^{3}\omega_{2f}(k)\tau'}
\nonumber\\
&&+ \beta_{k}e^{ia_{2f}^{3}\omega_{2f}(k)\tau'}\bigg],
\label{eq:alphabeta}
\end{eqnarray}
where  $\omega_{2f}(k) \equiv \sqrt{(k/a_{2f})^2 + m^2}$.
Here, one has
\begin{equation}
\label{a2b2}
\left|\alpha_{k}\right|^{2}-\left|\beta_{k}\right|^{2}=1,
\end{equation}
from the conserved Wronskian of Eq.~(\ref{eq:evot}).
The quantity
$\left|\beta_{k}\right|^{2}$ is the average number of particles in
mode-${\vec k}$ created by the expansion of the scale factor
from a state that initially has no particles \cite{Parker2,Parker3}.
Because our scale factor is asymptotically Minkowskian, the
meaning of particles at early and late times has no ambiguity.
The late-time creation and annihilation operators,
$a^{\dagger}_{\vec{k}}$ and $a_{\vec{k}}$, are related
to the early-time creation and annihilation operators
through a Bogoliubov transformation:
\begin{equation}
a_{\vec{k}}=\alpha_{k}A_{\vec{k}}+\beta_{k}^{*}A_{-\vec{k}}^{\dagger}.
\label{eq:Bogoliubov}
\end{equation}

\section{Particle Number in Pure and Mixed States}
\label{sec:number}

As noted above, if no particles are present at early times, then
the average particle number at late times in mode $\vec{k}$ is
\begin{equation}
\left<N_{\vec{k}}\right>_{t\rightarrow\infty}=\left<0\right|a^{\dagger}_{\vec{k}}a_{\vec{k}}\left|0\right>=\left|\beta_{k}\right|^{2},
\end{equation}
where $\left|0\right>$ is the state annihilated by the early-time
annihilation operators $A_{\vec{k}}$.

Let us define $\left|\delta\phi^{\rm (un)}_{k}\right|^{2} \equiv
\langle0\left|(\delta{\hat \phi}_{\vec k}){}^2\right|0\rangle_{\rm un}
= |f_{\vec{k}}|^2$.  Here,  ``un" refers to
unrenormalized values. In the continuum limit, this reduces to
$\left|\delta\phi^{\rm (un)}_{k}\right|^{2}=(2\pi)^{-3}|\psi_k|^2$.

Later, we will find that renormalization is necessary and
may have a significant effect on the magnitude of the
dispersion, even for modes that leave imprints on the
CMB and large-scale structure that can be observed
in the present universe. Let
$\left|\delta\phi^{\rm (re)}_{k}\right|^{2} \equiv
\langle0\left|(\delta{\hat \phi}_{\vec k}){}^2\right|0\rangle_{\rm re}$,
where
\begin{equation}
\sum_{\vec k} \,
\langle0\left|(\delta{\hat \phi}_{\vec k}){}^2\right|0\rangle_{\rm re}
\equiv  \langle0\left|(\delta{\phi}(x) ){}^2\right|0\rangle_{\rm re}.
\label{eq:defRenormModes}
\end{equation}
The value of
$\langle0\left|(\delta{\phi}(x) ){}^2\right|0\rangle_{\rm re}$
would diverge without renormalization.  In Minkowski space,
the renormalization would be equivalent to subtracting the vacuum
zero-point contributions. In the expanding universe there are
also divergent contributions coming from the time-dependence
of the scale factor $a(t)$, and to take those into account we
must use a curved spacetime renormalization method such
as adiabatic regularization or point-splitting Hadamard regularization.
We will take this up again later in Section \ref{sec:dens}

If the initial state of the universe were not the vacuum, but
instead were a statistical mixture of pure states,
each of which contains a definite number of particles at early
times, then we would find the analog of stimulated emission,
where the initial presence of scalar particles
tends to increase the number of scalar particles created by the
expansion of the universe \cite{Parker2, Parker3}:
\begin{equation}
\left<N_{\vec{k}}\right>_{t\rightarrow\infty}=
\left<N_{\vec k}{}^{0}\right>+\left|\beta_{k}\right|^{2}
\left(1+ 2\left<N_{\vec k}{}^{0}\right> \right).
\label{eq:stimulated-emission}
\end{equation}
Here, $\left<N_{\vec k}\right>_{t\rightarrow\infty}$
is the average particle number in mode $\vec{k}$
at late times, and $\left<N_{\vec k}{}^{0}\right>$
is the average number of particles in mode $\vec{k}$
at early times.   Because of this stimulated-emission
effect, the initial presence of particles could lead
to larger inflaton perturbations than would be the
case with an initial Minkowski vacuum state. Furthermore,
it could affect the scale invariance of the
inflaton perturbation spectrum at late times. The particles
that were present
at early times disperse so that their relic density becomes
negligible as a result of inflation. This is evident from the
first term on the right of Eq.~(\ref{eq:stimulated-emission}).
However, the density of quantized inflaton perturbations created
as a result of the inflationary expansion remains significant
and is determined by the second term.
In that term, the factor of
$\left|\beta_{k}\right|^{2}$
multiplying
$\left<N_{\vec k}{}^{0}\right>$
makes the effect of the initial number of particles present in the
coordinate volume $L^3$ significant
if $\left<N_{\vec k}{}^{0}\right>$
is larger than or of order $1$.
This effect could conceivably lead to observable consequences for
the large scale structure of the universe, and will be considered
in later work.  In the rest of this paper,
we assume our initial state is asymptotically a Minkowski vacuum.

\section{Joining Conditions for $\psi_{k}$}

\label{sec:match}

Consider a spacetime composed of three segments of the scale factor,
$a(t)$, in a homogeneous background metric given by
Eq.~(\ref{eq:metric}). For an example, see Figs.~\ref{fig:diagram}
and~\ref{fig:fluc}. The first and second segments are joined at the
time $t_{1}$, and the second and third segments are joined at the
time $t_{2}$.

We have two linearly independent solutions to the evolution equation
in both the second segment, with solutions $h_{1}(t)$ and
$h_{2}(t)$; and the third segment, with solutions $g_{1}(t)$ and
$g_{2}(t)$; for a total of four separate functions. These functions
are multiplied by constant coefficients that we must determine.
During the second segment, from $t_{1}$ to $t_{2}$, we have:
\begin{eqnarray}
\psi_{k}(t)=Ah_{1}(t)+Bh_{2}(t), \\
\nonumber \psi_{k}'(t)=Ah_{1}'(t)+Bh_{2}'(t).
\label{eq:ABcoeff}
\end{eqnarray}
For $t>t_{2}$, we have:
\begin{eqnarray}
\psi_{k}(t)=Cg_{1}(t)+Dg_{2}(t), \\
\nonumber \psi_{k}'(t)=Cg_{1}'(t)+Dg_{2}'(t).
\end{eqnarray}
If we require that $\psi_{k}(t)$ and $\psi_{k}'(t)$ be continuous at
$t_{1}$ and $t_{2}$. This imposes 4 matching conditions:
\begin{eqnarray}
Ah_{1}(t_{1})+Bh_{2}(t_{1})=\psi_{k}(t_{1}), \\
\nonumber Ah_{1}'(t_{1})+Bh_{2}'(t_{1})=\psi_{k}'(t_{1}), \\
\nonumber Cg_{1}(t_{2})+Dg_{2}(t_{2})=Ah_{1}(t_{2})+Bh_{2}(t_{2}), \\
\nonumber
Cg_{1}'(t_{2})+Dg_{2}'(t_{2})=Ah_{1}'(t_{2})+Bh_{2}'(t_{2}).
\end{eqnarray}
The calculation given in Appendix~\ref{appendix:A} then shows us
that
\begin{eqnarray}
C&=&\frac{1}{\left(g_{1}'g_{2}-g_{1}g_{2}'\right)_{t=t_{2}}}\times
\nonumber\\
&&\bigg\{\left[\frac{\psi_{k1}'h_{2}-\psi_{k1}
h_{2}'}{h_{1}'h_{2}-h_{1}h_{2}'}\right]_{t=t_{1}}\left(h_{1}'g_{2}-h_{1}g_{2}'\right)_{t=t_{2}}
\nonumber\\
&&+\left[\frac{\psi_{k1}'h_{1}-\psi_{k1}
h_{1}'}{h_{2}'h_{1}-h_{2}h_{1}'}\right]_{t=t_{1}}\left(h_{2}'g_{2}-h_{2}g_{2}'\right)_{t=t_{2}}\bigg\},
\nonumber\\
\
\end{eqnarray}
\begin{eqnarray}
D&=&\frac{1}{\left(g_{2}'g_{1}-g_{2}g_{1}'\right)_{t=t_{2}}}\times
\nonumber\\
&&\bigg\{\left[\frac{\psi_{k1}'h_{2}-\psi_{k1}
h_{2}'}{h_{1}'h_{2}-h_{1}h_{2}'}\right]_{t=t_{1}}\left(h_{1}'g_{1}-h_{1}g_{1}'\right)_{t=t_{2}}
\nonumber\\
&&+\left[\frac{\psi_{k1}'h_{1}-\psi_{k1}
h_{1}'}{h_{2}'h_{1}-h_{2}h_{1}'}\right]_{t=t_{1}}\left(h_{2}'g_{1}-h_{2}g_{1}'\right)_{t=t_{2}}\bigg\},
\nonumber\\
\
\end{eqnarray}
where $\psi_{k1}\equiv\psi_{k}(t_{1})$ and
$\psi_{k1}'\equiv\psi_{k}'(t_{1})$.  We find $\psi_{k1}$ and
$\psi_{k1}'$ from the solution to the evolution equation in the
initial asymptotically flat segment of the scale factor.  In the
massless case, this solution is given by
Eq.~(\ref{eq:hypergeometric}) below.  The functions $h_{1}(t)$ and
$h_{2}(t)$ are to be related to the evolution equation solutions in
the inflationary middle segment of the scale factor, which are given
in Eqs.~(\ref{eq:infleigen}) and~(\ref{eq:massinfleigen}) below. In
terms of those solutions, one finds that $A$ and $B$ in
Eq.~(\ref{eq:ABcoeff}) are given by $A=E(k)$ and $B=F(k)$.
Similarly, the functions $g_{1}(t)$ and $g_{2}(t)$ are related to
the solution of the evolution equation in the final asymptotically
flat segment of the scale factor.  The latter solution is given in
Eq.~(\ref{eq:hypergeometric2}), which allows us to specify that
$C=N_{1}(k)$ and $D=N_{2}(k)$.

\section{Massless Particle Production}
\label{sec:masslesspc}

We will
first consider the case, $m=0$.  Rewriting the evolution equation,
Eq.~(\ref{eq:evot}), in terms of $\tau$ instead of $t$ leads to
\begin{equation}
\frac{d^{2}\psi_{k}}{d\tau^2}=-k^{2}a^{4}\psi_{k}.
\label{eq:evotau}
\end{equation}

For the first segment of our composite scale factor, the solution of
(\ref{eq:evotau}) having positive frequency form
(\ref{eq:minkowski}) at early times is the hypergeometric function
\cite{ParkerToms,CinNature}
\begin{eqnarray}
\psi_{k}(t(\tau))&=&\frac{1}{\sqrt{2a_{1i}{}^{2}k}}
e^{-ika_{1i}{}^{2}\tau}
F(-ika_{1i}{}^{2}s_{i} +ika_{2i}{}^2 s_{i},
\nonumber\\
&&-ika_{1i}{}^{2}s_{i}-ika_{2i}{}^{2}s_{i};1-2ika_{1i}{}^2
s_{i};-e^{\frac{\tau}{s_{i}}}),
\nonumber\\
\
\label{eq:hypergeometric}
\end{eqnarray}
where $F(a,b;c;d)$ is the hypergeometric function as defined
in \cite[see 15.1.1]{AbramowitzStegun}.

For the exponentially expanding segment of the scale factor in the
massless case
\begin{eqnarray}
\psi_{k}(t)=&-a(t)^{-\frac{3}{2}}&\frac{i}{2}\
\sqrt{\frac{\pi}{H_{\rm infl}}}\
\bigg[E(k)H_{\frac{3}{2}}^{(1)}\left(\frac{k}{a(t)H_{\rm
infl}}\right)
\nonumber\\
&&+\ F(k)H_{\frac{3}{2}}^{(2)}\left(\frac{k}{a(t)H_{\rm
infl}}\right)\bigg], \label{eq:infleigen}
\end{eqnarray}
where $H^{(1)}$ and $H^{(2)}$ are the Hankel functions of the first
and second kind.  The variables $t$ and $\tau$ are related by
Eq.~(\ref{eq:tvtau}). The coefficients $E(k)$ and $F(k)$
are determined by the matching conditions of the first joining point
at $t=t_{1}$. We note that the finite period of exponential
inflation lacks the full symmetries of a de Sitter universe. In the
pure de Sitter case, as shown in \cite{AllenFolacci}, the $k=0$ mode
has to be chosen in a special way to avoid infrared divergences. For
our $a(t)$, infrared divergences do not arise (see
Sec.~\ref{sec:masslesspc}).

For the final segment of our composite scale factor,
the solution of the evolution equation (\ref{eq:evotau}) is a linear
combination of hypergeometric functions
\cite{ParkerToms,CinNature}:
\begin{eqnarray}
\psi_{k}(t(\tau'))
&=&N_{1}(k)e^{-ika_{1f}{}^{2}\tau'}F(-ika_{1f}{}^{2}s_{f}
+ika_{2f}{}^{2} s_{f},
\nonumber\\
&&-ika_{1f}{}^{2}s_{f}-ika_{2f}{}^{2}s_{f};1-2ika_{1f}{}^{2}
s_{f};-e^{\frac{\tau'}{s_{f}}})
\nonumber\\
&&+N_{2}(k)e^{ika_{1f}{}^{2}\tau'}F(ika_{1f}{}^{2}s_{f}
+ika_{2f}{}^{2}
s_{f},
\nonumber\\
&&ika_{1f}{}^{2}s-ika_{2f}{}^{2}s_{f};1+2ika_{1f}{}^{2}
s_{f};-e^{\frac{\tau'}{s_{f}}}),
\nonumber\\
\ \label{eq:hypergeometric2}
\end{eqnarray}
where the coefficients $N_{1}(k)$ and $N_{2}(k)$ are determined by
the matching conditions of the second joining point at $t=t_{2}$.

\begin{figure}[hbtp]
\includegraphics[scale=1.4]{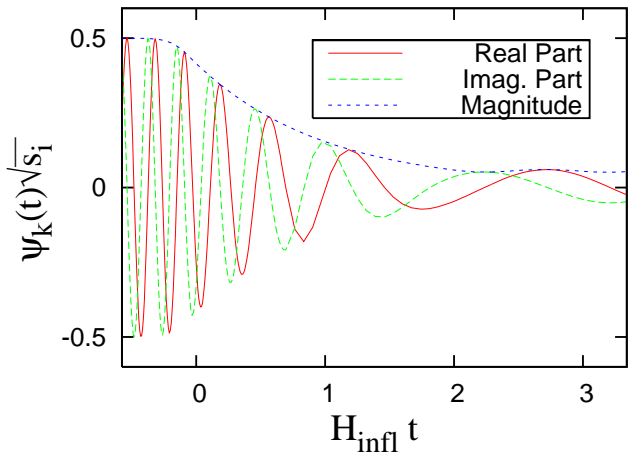}
\caption{\label{fig:fluc} (color online). A solution
to the evolution equation for a particular Fourier mode
$\psi_{k}(t)$
is plotted
versus dimensionless time for the same composite scale factor
$a(t)$ shown in Fig.~\ref{fig:diagram}.
The real and imaginary parts of
the dimensionless quantity $\sqrt{s_{i}}\psi_{k}(t)$,
and its magnitude are all plotted for $k=2$.}
\end{figure}

An example of the evolution for a particular mode is plotted for a
specific choice of parameters using our composite scale factor in
Fig.~\ref{fig:fluc}.  We emphasize that our considerations below
apply to any length of the middle exponentially inflating segment
of the expansion of the universe, not just to the one
used in Figs.~\ref{fig:diagram} and~\ref{fig:fluc}.

For our choice of the final asymptotically flat segment given by
Eq.~(\ref{eq:aHG}), where we use Eq.~(\ref{eq:hypergeometric2}) to
define our functions $g_{1}(t)$ and $g_{2}(t)$ in terms of the
relationship $\psi_{k}(t)=N_{1}g_{1}(t(\tau))+N_{2}g_{2}(t(\tau))$,
we find the coefficients $\alpha_{k}$ and $\beta_{k}$ of
Eq.~(\ref{eq:alphabeta}) from the large argument asymptotic forms
\cite{CinNature,ParkerToms,AbramowitzStegun}. With $b_{f}=0$,
$c_{1}\equiv iks_{f}a_{1f}^{\ \ 2}$, and $c_{2}\equiv
iks_{f}a_{2f}^{\ \ 2}$, we have

\begin{eqnarray}
\alpha_{k}=&&\sqrt{2ka_{2f}^{\ \ 2}}\bigg[\frac{C\
\Gamma(1-2c_{1})\Gamma(-2c_{2})}{\Gamma(1-c_{1}-c_{2})\Gamma(-c_{1}-c_{2})}
\nonumber\\
&&+\frac{D\
\Gamma(1+2c_{1})\Gamma(-2c_{2})}{\Gamma(1+c_{1}-c_{2})\Gamma(c_{1}-c_{2})}\bigg],
\label{eq:alpha}
\end{eqnarray}
and
\begin{eqnarray}
\beta_{k}=&&\sqrt{2ka_{2f}^{\ \ 2}}\bigg[\frac{C\
\Gamma(1-2c_{1})\Gamma(2c_{2})}{\Gamma(1-c_{1}+c_{2})\Gamma(-c_{1}+c_{2})}
\nonumber\\
&&+\frac{D\
\Gamma(1+2c_{1})\Gamma(2c_{2})}{\Gamma(1+c_{1}+c_{2})\Gamma(c_{1}+c_{2})}\bigg].
\label{eq:beta}
\end{eqnarray}
Recall that $C$ and $D$ and the functions $g_1(t)$ and $g_2(t)$
were defined in Sec.~\ref{sec:match}.  As the Wronskian of the
evolution equation is conserved with our joining conditions,
Eq.~(\ref{a2b2}) should hold to arbitrary accuracy.
As a useful check on our calculations, we verified that in
all the cases considered in this paper, the relation
$\left|\alpha_{k}\right|^{2}-\left|\beta_{k}\right|^{2}=1$
is satisfied to at least 500 significant figures.

The quantity $|\beta_{k}|^2$ is the average number of particles
present in mode k at late times. These particles are created
by the expansion of the universe through $N_{e}$ e-folds
from a state having no particles present at early times.

\begin{figure}[hbtp]
\includegraphics[scale=1.4]{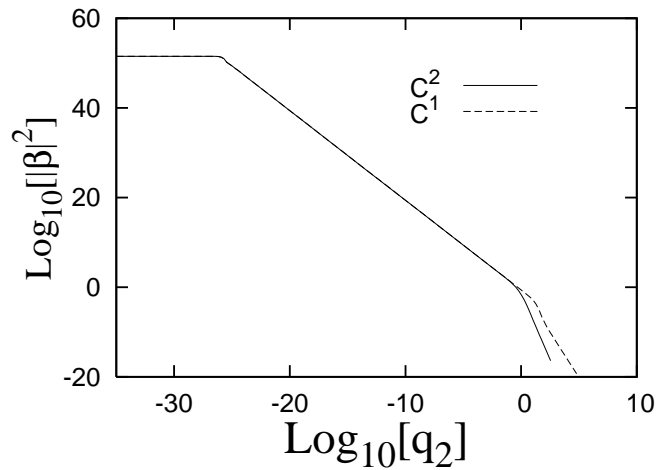}
\caption{\label{fig:b2vq2} Average late time particle number per
mode, ($|\beta_{q_{2}}|^{2}$), versus $q_{2} = k/(a_{2f}H_{\rm
infl})$ for 60 e-folds of inflation. Two cases are plotted for the
massless case based on the behavior at the matching conditions: the
scale factor continuous in 0th, 1st, and 2nd derivatives ($C^{2}$);
and the scale factor continuous in 0th and 1st derivatives
($C^{1}$). Note that in the $C^{1}$ case,
$\left|\beta_{q_{2}}\right|^{2}$ transitions from a $q_{2}^{\ -2}$
dependence at the end of the intermediate-$q_{2}$ region all the way
to a $q_{2}^{\ -6}$ dependence, temporarily parallel to the $C^{2}$
large-$q_{2}$ regime, before settling down into its ultraviolet
$q_{2}^{\ -4}$ behavior. For the wiggles near the transition from
the small-$q_{2}$ region to the intermediate-$q_{2}$ region at
$q_{2}=e^{-N_{e}}$, compare with the graph of the dispersion
spectrum in Fig.~\ref{fig:2pt}.}
\end{figure}

We use the dimensionless variable
\begin{equation}
q_{2}\equiv\frac{k}{a_{2f}H_{\rm infl}}, \label{q2}
\end{equation}
where k is the wave number, $a_{2f}$ is the constant scale factor
that is approached at very late times, and $H_{\rm infl}$ is the constant value of $\left({\dot{a}(t)}/{a(t)}\right)$ during the exponential
expansion of the middle segment. It is convenient to express our
results in terms of the dimensionless quantity $q_2$.  For example,
for $N_e = 60$, as in Fig.~\ref{fig:b2vq2}, the graph is the
same if we change the range of $k$, and the values of $a_{2f}$
and $H_{\rm infl}$, while keeping the range of $q_2$ unchanged.
We will often write
$\left|\beta_{q_{2}}\right|^{2}$
for the quantity
$\left|\beta_{k}\right|^{2}$ with $k = q_{2} a_{2f} H_{\rm infl}$.

We define three regions of $q_{2}$.  Values of
$q_{2}\lesssim\exp(-N_{e})$ are in the ``small-$q_{2}$ region."
Values of $\exp(-N_{e})\lesssim q_{2}\lesssim1$ are in the
``intermediate-$q_{2}$ region."  Values of $1\lesssim q_{2}$ are in
the ``large-$q_{2}$ region."

When $a(t)$ is at least $C^{1}$, i.e. when $H_{\rm infl}$ is
continuous, we find numerically that the particle production per
mode in the small-$q_{2}$ region, ($q_{2}\lesssim e^{-N_{e}}$), is
to good approximation given by
\begin{equation}
\beta_{q_{2}} = \sinh[N_{e}].
\label{eq:smallqbeta}
\end{equation}
We also find this to be the case,
analytically, by taking the limit $k\rightarrow0$.  For at least a
moderate number of e-folds, this simplifies to
\begin{equation}
|\beta_{q_{2}}|^{2}\simeq\frac{1}{4}e^{2N_{e}}.
\label{eq:p21}
\end{equation}
The dependence in the intermediate-$q_{2}$ region
($e^{-N_{e}}\lesssim q_{2}\lesssim 1$) for the $C^{2}$ or $C^{1}$
massless case is
\begin{equation}
|\beta_{q_{2}}|^{2}\simeq\frac{1}{4}q_{2}^{\ -2}.
\label{eq:p22}
\end{equation}

When $N_{e}$ is finite, with our composite scale factor there are no
infrared divergences.  For infinite inflation, where
$N_{e}\rightarrow\infty$, we find the infrared divergences of a de
Sitter universe.  This problem is resolved for a true de Sitter
universe in \cite{AllenFolacci}.  Our composite scale factor is
different from a purely de Sitter universe in that our initial
conditions are specified by our initial asymptotically flat region
of the scale factor.

Discontinuities in the derivatives of the scale factor at the
matching points cause increased particle production for
large-$q_{2}$, (i.e., $q_{2}\gtrsim1$). This is evident in
Fig.~\ref{fig:b2vq2}.  As shown by Parker in \cite{Parker2,
Parker3}, the particle number created by the expansion of the
universe is related to an adiabatic invariant of the harmonic
oscillator, and for such an oscillator the continuity of the
frequency and its derivatives is related to the change in the
adiabatic invariant \cite{Kulsrud}. The increased particle
production associated with discontinuities in the derivatives of the
scale factor $a(t)$ is a consequence of this relation. This result
is also seen in \cite{Chung}.

 For the
$C^{1}$ case, where the scale factor and $H=\dot{a}(t)/a(t)$ are
both continuous, the large-$q_{2}$ region goes like
\begin{equation}
|\beta_{q_{2}}|^2=n_{4}q_{2}^{-4}.
\end{equation}
For the $C^{2}$ case, where the scale factor and $H=\dot{a}(t)/a(t)$
and $\dot{H}(t)$ are all continuous, the large-$q_{2}$ region goes
like
\begin{equation}
|\beta_{q_{2}}|^2=n_{6}q_{2}^{-6}.
\label{eq:rapidFall}
\end{equation}
Here $n_{4}$ and $n_{6}$ are constant coefficients, with
$n_{4}\simeq n_{6}\simeq\mathcal{O}(1)$ for a gradual end to
inflation.  For a sufficiently abrupt end to inflation, $n_{4}$ and
$n_{6}$ can be made to be arbitrarily large. See
Sec.~\ref{sec:reheat}.

When $H(t)$ is not continuous, we find quite a different behavior in
the $C^{0}$ case.  The evolution equation, Eq.~(\ref{eq:evot}), may
be written \cite{Parker3}
\begin{equation}
\frac{d^{2}\psi_{k}(t)}{dt^{2}}+\left[\frac{k^{2}}{a(t)^{2}}+m^{2}-\frac{3}{4}\left(\frac{\dot{a}(t)}{a(t)}\right)^{2}-\frac{3}{2}\frac{\ddot{a}(t)}{a(t)}\right]\psi_{k}(t)=0.
\end{equation}
At the discontinuity in $\dot{a}(t)$ if we express the jump as a
step function, then the form of $\ddot{a}(t)$ picks up a
delta-function contribution. Thus, there is a finite jump in
$d\psi_{k}(t)/dt$ across the discontinuity. In the $C^{0}$ case,
$\left|\beta_{q_{2}}\right|^{2}$ is proportional to $q_{2}^{\ -2}$
in the small- and large-$q_{2}$ regions, and it is proportional to
$q_{2}^{\ -4}$ in the intermediate-$q_{2}$ region. A $C^{0}$
scenario would suffer from these problems in addition to the
divergences mentioned earlier, hence we will not consider it
further.

For a non-composite scale factor composed of one asymptotically flat
scale factor defined by Eq.~(\ref{eq:aHG}), for large values of
$q_{2}$ the value of $|\beta_{q_{2}}|^{2}$ falls off faster than any
power of $q_{2}$, and in terms of $k$ we have:
\cite{CinNature,ParkerToms}
\begin{eqnarray}
&&|\beta_{k}|^2=
\nonumber\\
&&\frac{\sin^{2}\left(\frac{1}{2}[1-\sqrt{1+4k^{2}s^{2}b}]\right)+\sinh^{2}[\pi
ks(a_{1}^{\ 2}-a_{2}^{\ 2})]}{\sinh^{2}[\pi ks(a_{1}^{\ 2}+a_{2}^{\
2})]-\sinh^{2}[\pi ks(a_{1}^{\ 2}-a_{2}^{\ 2})]}.
\nonumber\\
&&\
\end{eqnarray}
In the limit that $k\rightarrow0$ for the case of the scale factor
of Eq.~(\ref{eq:aHG}), which is asymptotically flat at early and
late times and has no exponential segment, we find that
$\lim_{k\rightarrow0}\left|\beta_{k}\right|^{2}=\sinh^2[N_{e}]$,
where in this case $N_{e}$ is $\ln\left(a_{2}/a_{1}\right)$. This is
the same small-$q_{2}$ limit for the average number of particles
created per mode as we found above in Eq.~(\ref{eq:smallqbeta}).

\section{Dispersion Spectrum}
\label{sec:MasslessDispersionSpectrum}

The unrenormalized dispersion spectrum
is \cite{ParkerToms,ParkerNY}
\begin{equation}
\left< 0 \right|\delta\phi^{2}\left| 0 \right>_{\rm un}
=\frac{1}{2(a_{2f}L)^{3}}\sum_{k}\left[\frac{1+2|\beta_{k}|^{2}}{\sqrt{(k/a_{2f})^{2}+m^{2}}}\right],
\label{eq:2pt}
\end{equation}
where the expectation value is with respect to the state
$\left|0\right>$ having no particles at early times.

\begin{figure}[hbtp]
\includegraphics[scale=1.4]{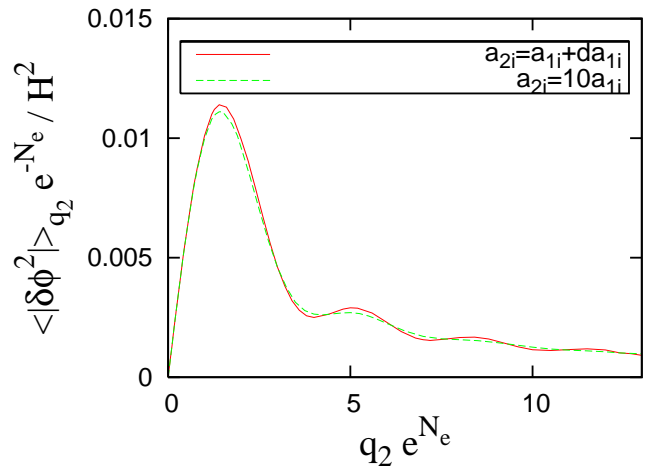}
\caption{\label{fig:2pt} (color online).  Dispersion $\left<\
\right|\delta\phi^{2}\left|\ \right>_{q_2}/H_{\rm infl}^{2}$ given
by Eq.~(\ref{eq:disp}) for our composite scale factor continuous in
$a(t)$, $\dot{a}(t)$, and $\ddot{a}(t)$ over an expansion of 60
e-folds.  The y-axis, $\left<\ \right|\delta\phi^{2}\left|\
\right>_{q_2}/H_{\rm infl}^{2}$, is shown multiplied by a factor of
$e^{-N_{e}}$; and the x-axis, $q_{2}$, is shown multiplied by a
factor of $e^{N_{e}}$. When using this scaling, for a given set of
values of $a_{1i}$ and $a_{2i}$ the region plotted
in this graph would look identical for any number of
e-folds larger than about 10. In the
case of $a_{2i}=a_{1i}+da_{1i}$, where
$da_{1i}\equiv10^{-26}a_{1i}$, we see marked peaks in the dispersion
spectrum. When we change the parameters in the initial
asymptotically flat region to $a_{2i}=10a_{1i}$, these peaks are
somewhat damped, as shown.  The ending conditions of the final asymptotically flat segment do not affect these peaks.}
\end{figure}

\begin{figure}[hbtp]
\includegraphics[scale=1.4]{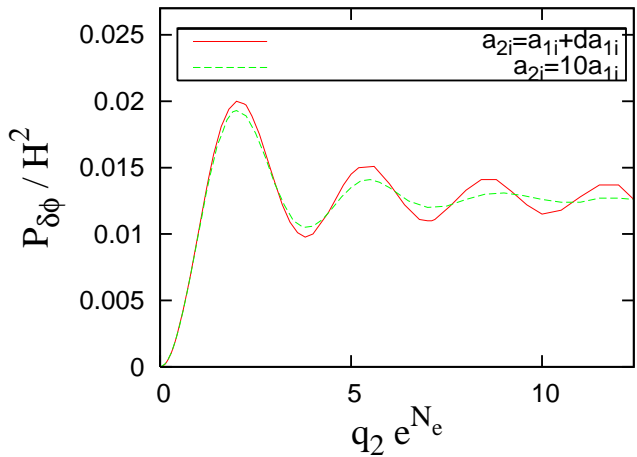}
\caption{\label{fig:dispSpec} (color online).  Dispersion spectrum
$\mathcal{P}_{\delta\phi}/H_{\rm infl}^{2}$ given by
Eq.~(\ref{eq:dispspec}) for our composite scale factor continuous in
$a(t)$, $\dot{a}(t)$, and $\ddot{a}(t)$ over an expansion of 60
e-folds.  The two cases considered are the same as explained
in Fig.~\ref{fig:2pt}.  Similar peaks were found in a function related
to the energy-momentum tensor in  \cite[see their Fig.~1]{Anderson}.}
\end{figure}

\begin{figure}[hbtp]
\includegraphics[scale=1.4]{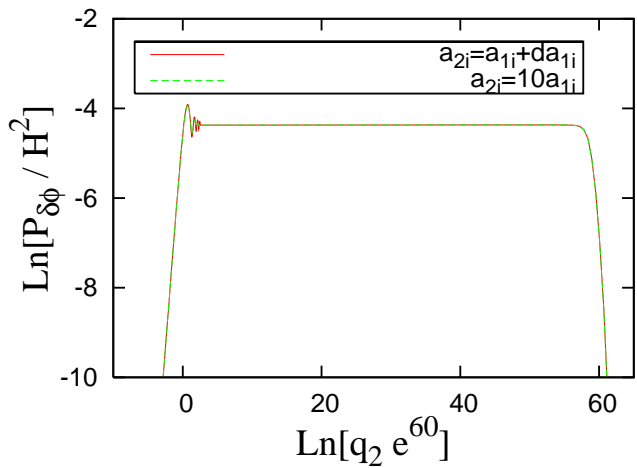}
\caption{\label{fig:lnDispSpec} (color online).  Dispersion
spectrum $\mathcal{P}_{\delta\phi}/H_{\rm infl}^{2}$ given by
Eq.~(\ref{eq:dispspec}) for our composite scale factor continuous in
$a(t)$, $\dot{a}(t)$, and $\ddot{a}(t)$ over an expansion of 60
e-folds.  The cases considered are the same as those in
Figs.~\ref{fig:2pt} and \ref{fig:dispSpec}. Here we plot the data
from Fig.~\ref{fig:dispSpec} on a $\ln$-$\ln$ scale over a wider
range of $q_{2}$ for a gradual end to inflation, where
$\ln(a_{2f}/a_{1f})\simeq1$.}
\end{figure}

We will first consider the massless case, $m=0$.  See
Sec.~\ref{sec:mass} for the massive case.  If
$\left|\beta_k\right|^{2}$ were 0 in Eq.~(\ref{eq:2pt}), then one would be left with the contribution of the late-time Minkowski space vacuum.
As is standard in flat spacetime quantum field theory,
we subtract off this contribution of the Minkowski vacuum,
leaving the physically relevant, renormalized dispersion,
\begin{eqnarray}
\left< 0 \right|\delta\phi^{2}\left| 0 \right>_{\rm re}
&&=\frac{1}{2(a_{2f}L)^{3}}\sum_{k}\frac{2|\beta_{k}|^{2}}{\sqrt{(k/a_{2f})^{2}}}
\nonumber\\
&&=\frac{1}{a_{2f}^{\ \ 2}L^{3}}\sum_{k}\frac{|\beta_{k}|^{2}}{k},
\label{eq:MinkSubtraction}
\end{eqnarray}
where the subscript ``re" refers to renormalized.
In the continuum limit, this becomes
\begin{equation}
\left< 0 \right|\delta\phi^{2}\left| 0 \right>_{\rm
re}=\frac{1}{a_{2f}^{\ \
2}(2\pi)^{3}}\int_{0}^{\infty}\frac{|\beta_{k}|^{2}}{k}d^{3}k,
\end{equation}
or with spherical symmetry,
\begin{equation}
\left< 0 \right|\delta\phi^{2}\left| 0 \right>_{\rm
re}=\frac{1}{2\pi^{2}a_{2f}^{\ \
2}}\int_{0}^{\infty}k|\beta_{k}|^{2}dk.
\end{equation}
With $k=q_{2}a_{2f}H_{\rm infl}$ and $dk=dq_{2}a_{2f}H_{\rm infl}$,
we have
\begin{equation}
\left< 0 \right|\delta\phi^{2}\left| 0 \right>_{\rm re}=\frac{H_{\rm
infl}^{2}}{2\pi^{2}}\int_{0}^{\infty}q_{2}|\beta_{q_{2}}|^{2}dq_{2}.
\end{equation}
The $q_2$-component of the dispersion in the massless case is thus,
\begin{equation}
\left<\ \right|\delta\phi^{2}\left|\ \right>_{q_2}
\equiv\frac{q_{2}|\beta_{q_{2}}|^{2}H_{\rm infl}^{2}}{2\pi^{2}}.
\label{eq:disp}
\end{equation}

In the late-time Minkowski space, this subtraction is all that is
necessary, but during the time when the universe is expanding
this subtraction alone would give an infinite dispersion when
summed over all modes.   During the expansion, there are
additional subtractions necessary.
Adiabatic regularization \cite{Parker5} includes those
and reduces to this standard Minkowski space subtraction
in flat spacetime.  In Sec.~\ref{sec:dens}, we consider
the scalar perturbations of the metric that are
created  by the inflaton fluctuations. Those metric
perturbations for a given mode are formed during
inflation shortly after the inflaton mode has exited the
Hubble sphere.  As explained in \cite{Parker1, Agullo2},
we regard the metric perturbations as classical at
the time when they are produced and assume that they
respond to the {\em renormalized} dispersion of the inflaton
fluctuation field. This implies that the relevant adiabatic
subtractions that influence the metric perturbation for
a given mode are those evaluated near the time that
the quantized inflaton fluctuations exit the Hubble sphere.

The definition of the spectrum of inflaton perturbations,
$\mathcal{P}_{\delta\phi}$, given in \cite{LL} is
\begin{equation}
\left< 0 \right|\delta\phi^{2}\left| 0 \right>_{\rm
re}=\int_{0}^{\infty}\mathcal{P}_{\delta\phi}\frac{dq_{2}}{q_{2}}.
\label{eq:defp}
\end{equation}
Therefore, we obtain the spectrum of inflaton fluctuations as
\begin{equation}
\mathcal{P}_{\delta\phi}=
\frac{q_{2}^2|\beta_{q_{2}}|^{2}H_{\rm infl}^{2}}{2\pi^{2}}.
\label{eq:dispspec}
\end{equation}

We plot $\left<\ \right|\delta\phi^{2}\left|\ \right>_{q_2}/H_{\rm
infl}^{2}$ in Fig.~\ref{fig:2pt}, and we plot
$\mathcal{P}_{\delta\phi}/H_{\rm infl}^{2}$ in
Fig.~\ref{fig:dispSpec}. The peaks in these figures do not depend on
the number of e-folds and are the result of the state $| 0 \rangle$
having no particles in the asymptotically flat part of the expansion
at early times. In Fig.~\ref{fig:lnDispSpec}, we plot
$\ln({P}_{\delta\phi}/H_{\rm infl}^{2})$ versus $\ln(q_2 \exp(60))$
for an expansion of 60 e-folds. In this figure, we have plotted the
full range of the inflationary part of the expansion. We see that in
both of the cases plotted, where $a(t)$, $\dot{a}(t)$, and
$\ddot{a}(t)$ are all continuous; and the case where only $a(t)$ and
$\dot{a}(t)$ are continuous; $\left< 0 \right|\delta\phi^{2}\left| 0
\right>_{\rm re}$ is finite without the need for any renormalization
beyond subtracting off the Minkowski vacuum contribution at late
times.  If the first derivative of the scale factor were not
continuous, then the integrated dispersion would diverge.

\section{Massive Particle Production}
\label{sec:massivepc}

In the case of a massive scalar field, the evolution equation
(\ref{eq:evot}) with $a(t)$ given by the inflationary
exponential of Eq.~(\ref{eq:infl}), has the exact solution
\begin{eqnarray}
\psi_{k}(t)=&-a(t)^{-\frac{3}{2}}&\frac{i}{2}\sqrt{\frac{\pi}{H_{\rm
infl}}}\bigg[E(k)H_{\sqrt{\frac{9}{4}-m_{H}^{\
2}}}^{(1)}\left(\frac{k}{a(t)H_{\rm infl}}\right)
\nonumber\\
&&+F(k)H_{\sqrt{\frac{9}{4}-m_{H}^{\
2}}}^{(2)}\left(\frac{k}{a(t)H_{\rm infl}}\right)\bigg],
\label{eq:massinfleigen}
\end{eqnarray}
where
\begin{equation}
m_{H}\equiv\frac{m}{H_{\rm infl}}.
\end{equation}

\begin{figure}[hbtp]
\includegraphics[scale=1.4]{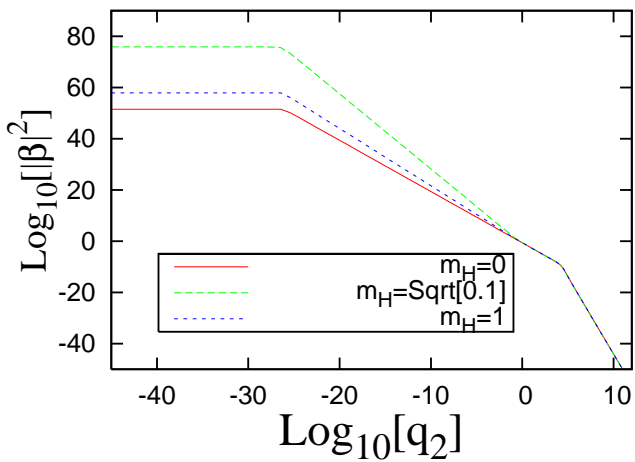}
\caption{\label{fig:massive} (color online).  Log-Log plot of
the average number of created particles versus $q_2$.
The effect of three different masses is
shown for an expansion of 60 e-folds.
The beginning and end segments of $a(t)$ are rather
abrupt, with $a_{2i}=a_{1i}(1+10^{-26})$,
$a_{2f}=a_{1i}e^{60}$, and $a_{1f}=0.9999a_{2f}$.
Comparing the graph
plotted here for $m_H = 0$ with the
corresponding graph plotted
for a gradual end to inflation
in Fig.~\ref{fig:b2vq2}, one sees that the two graphs
are the same for $q_{2}\lesssim1$.
In the present
graph, however, there is a ``stretched" region of
$\left|\beta_{q_{2}}\right|^{2}\propto q_{2}^{-2}$ to the right of
$q_{2}\simeq1$ that lasts until $q_{2}\simeq10^{4}$ before the
ultraviolet behavior of $\left|\beta_{q_{2}}\right|^{2}\propto
q_{2}^{-6}$ is seen.  (For more discussion, see
Sec.~\ref{sec:reheat}.) This stretching is due to particle
creation caused by the rapid change of $a(t)$ from
its value at the end of inflation to its final value of $a_{2f}$.
Both of
the massive cases shown here produce more particles of
low momentum than does the massless case.
See also Fig.~\ref{fig:domterm}.}
\end{figure}

\begin{figure}[hbtp]
\includegraphics[scale=1.4]{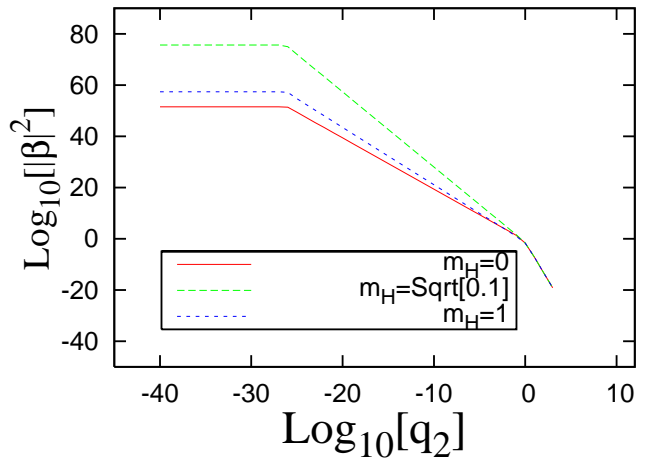}
\caption{\label{fig:domterm} (color online).  The dependence of
particle production ($\left|\beta_{q_{2}}\right|^{2}$) on mass is
shown for an expansion of 60 e-folds.  This graph is different from
Fig.~\ref{fig:massive} in that the transition from exponential
expansion to the final asymptotic segment of the scale factor is
more gradual, happening over about an e-fold.
For values of $q_{2}\lesssim1$, this graph is
identical to that of Fig.~\ref{fig:massive}.}
\end{figure}

The evolution equation for arbitrary $a(t)$
can be written in terms of $\tau$ (defined
in Eq.~(\ref{eq:tvtau})) as
\begin{equation}
\frac{d^{2}\psi_{k}}{d\tau^2}=-(k^{2}a^{4}+m^{2}a^{6})\psi_{k}.
\end{equation}
In the initial and final asymptotically flat
segments of $a(t)$ that are joined to
the inflationary segment at early and late times, we do not
have an exact solution of this equation for nonzero mass.
Therefore, we used approximations to obtain the plots in
Fig.~\ref{fig:massive} and Fig.~\ref{fig:domterm}.
In these graphs, we see that at low momentum there
are many more particles with $m_H = 0.1$ than with
$m_H = 1$ present at late times.

\section{Massive Dispersion Spectrum}

\label{sec:mass}

A calculation of the dispersion spectrum in the massive case leads
to an equation analogous to Eq.~(\ref{eq:disp}):
\begin{equation}
\left<\ \right|\delta\phi^{2}\left|\
\right>_{q_2}\equiv\frac{q_{2}|\beta_{q_{2}}|^{2}H_{\rm
infl}^{2}}{2\pi^{2}\sqrt{1+\frac{m_{H}^{\ 2}}{q_{2}^{\ 2}}}},
\label{eq:massdisp}
\end{equation}
where
\begin{equation}
\mathcal{P}_{\delta\phi}=q_{2}\left<\ \right|\delta\phi^{2}\left|\
\right>_{q_2}.
\end{equation}

\begin{figure}[hbtp]
\includegraphics[scale=1.4]{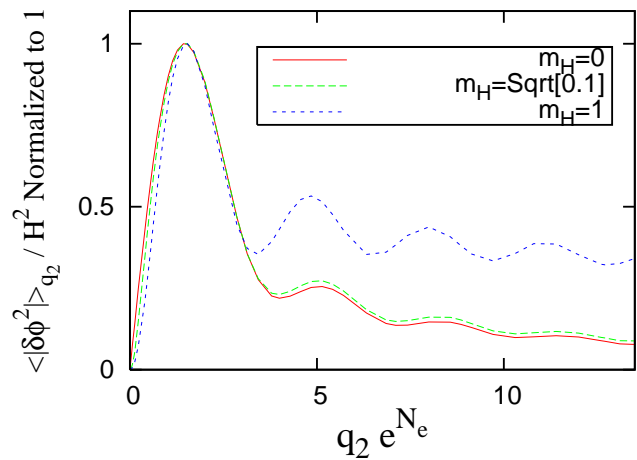}
\caption{\label{fig:massive2pt} (color online).  Comparison of
dispersion spectrum, $\left<\ \right|\delta\phi^{2}\left|\
\right>_{q_2}/H_{\rm infl}^{2}$, given by Eq.~(\ref{eq:massdisp})
and normalized to 1, for our composite scale factor continuous in
$a(t)$, $\dot{a}(t)$, and $\ddot{a}(t)$ over an expansion of 60
e-folds for various masses.  The values of $\left<\
\right|\delta\phi^{2}\left|\ \right>_{q_2}/H_{\rm infl}^{2}$ were
divided by the maximum value of the primary peak located at
$q_{2}\simeq\exp(-N_{3})$ for each. To normalize these peaks,
$\left<\ \right|\delta\phi^{2}\left|\ \right>_{q_2}/H_{\rm
infl}^{2}$ was divided by the following factors: $1.3\times10^{24}$
for the massless case, $2.3\times10^{22}$ for $m_{H}^{\ 2}=0.1$, and
$2.2\times10^{4}$ for $m_{H}^{\ 2}=1$.}
\end{figure}

\begin{figure}[hbtp]
\includegraphics[scale=1.4]{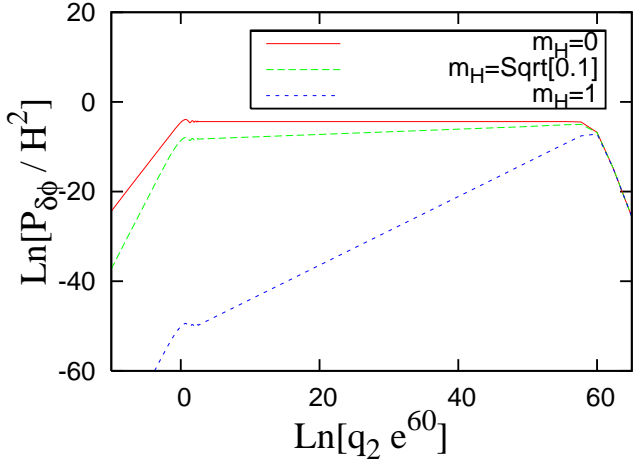}
\caption{\label{fig:lnMassDispSpec} (color online).  Dispersion
spectrum $\mathcal{P}_{\delta\phi}/H_{\rm infl}^{2}$ given by
Eq.~(\ref{eq:dispspec}) for our composite scale factor continuous in
$a(t)$, $\dot{a}(t)$, and $\ddot{a}(t)$ over an expansion of 60
e-folds.  The masses considered are the same as in
Figs.~\ref{fig:massive} and~\ref{fig:domterm}, where the average
number of created particles was plotted.}
\end{figure}

The dispersion spectrum is plotted for three different cases of
$m_{H}$ in Fig.~\ref{fig:massive2pt}. We have found that, even in
the massive case, the observed humps are dependent only upon the
initial conditions.  The shape of the curves is fixed above a
moderate number of e-folds.  We define the variable $J$, such that
the maximum value of $\left<\ \right|\delta\phi^{2}\left|\
\right>_{q_2}/H_{\rm infl}^{2}$ for the major peak, which is the
peak located nearest to $q_{2}=e^{-N_{e}}$, is $J\ e^{(P-1)N_{e}}$
in the massless case and is $J\ e^{(P-2)N_{e}}$ in the massive case.
Then, the normalization factor scales like $e^{(P-1)N_{e}}$ in the
massless case, as can be seen from Eq.~(\ref{eq:disp}); and the
normalization factor scales like $e^{(P-2)N_{e}}$ in the massive
case, as can be seen from Eq.~(\ref{eq:massdisp}), where we define
the exponent $P$ in the following way:
\begin{equation}
\left|\beta_{q_{2}}\right|^{2}\simeq\frac{1}{4}q_{2}^{\ -P}
\label{eq:P}
\end{equation}
in the region of intermediate-$q_{2}$, ($e^{-N_{e}}\lesssim
q_{2}\lesssim1$), and
\begin{equation}
\left|\beta_{q_{2}}\right|^{2}\simeq\frac{1}{4}e^{PN_{e}}
\end{equation}
in the small-$q_{2}$ region, ($q_{2}\lesssim e^{-N_{e}}$). The
exponent $P$ is well described by a $q_{2}$-independent value in the
case of $m=0$ and in the case of $0.01\lesssim m_{H}^{\
2}\lesssim9/4$.

\begin{figure}[hbtp]
\includegraphics[scale=1.4]{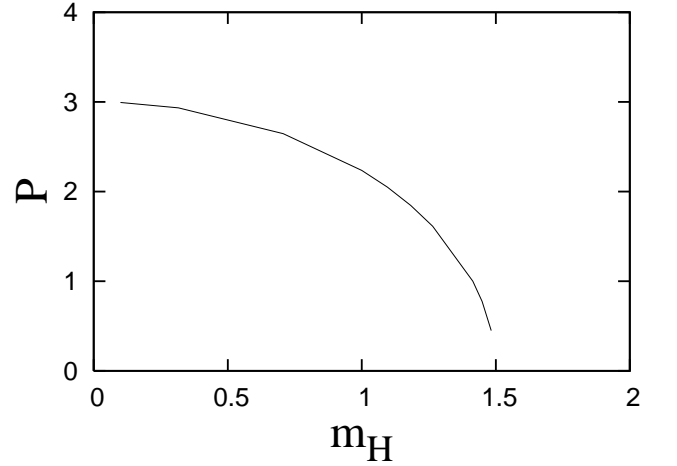}
\caption{\label{fig:pvmh} The dependence of the variable $P$, as
defined in Eq.~(\ref{eq:P}), upon $m_{H}=m/H_{\rm infl}$. The
calculated data points shown lie on the curve $P=\sqrt{9-4m_{H}^{\
2}}$.  Outside of the region plotted, however, $P$ does not have a
constant, $q_{2}$-independent value. For $m_{H}>1.5$, the argument,
$\sqrt{(9/4)-m_{H}^{\ 2}}$, of the Hankel functions becomes
imaginary, and $\left|\beta_{q_{2}}\right|^{2}$ oscillates with
changing $q_{2}$. For an example of a non-zero mass much smaller
than $H_{\rm infl}$, see Fig.~\ref{fig:tinymass}.}
\end{figure}

\begin{figure}[hbtp]
\includegraphics[scale=1.4]{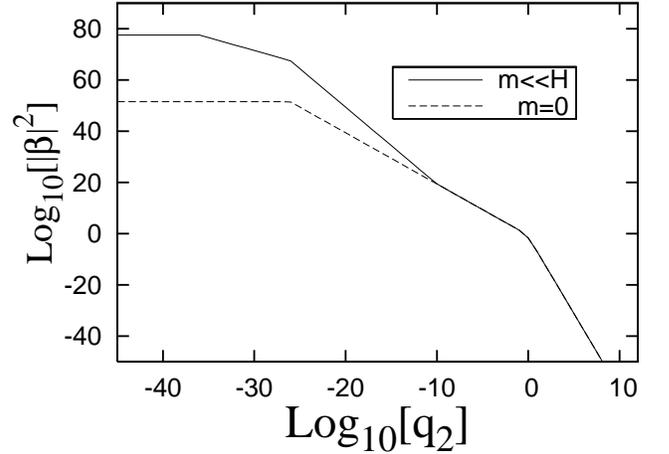}
\caption{\label{fig:tinymass} Particle production as a function of
$q_{2}$ is plotted for 60 e-folds for both the massless case and the
case of $m=10^{-10}H_{\rm infl}$, labeled as ``$m<<H$.'' As the mass
decreases, the region of overlap of the solid curve with the dotted
($m=0$) curve becomes larger and approaches the dotted curve in the
limit as $m\rightarrow 0$.  The solid curve breaks away from the
dotted curve when $q_2 \lesssim m_H$. The solid curve takes the
constant value, $(1/4)q_{2}^{\ 3N_{e}}$, when
$q_{2}<m_{H}\exp(-N_{e})$. In the region of
$m_{H}\exp(-N_{e})<q_{2}<m_{H}$, we see $(k/a(t))^{2}\gg m^{2}$ in
the initial asymptotically flat region and $(k/a(t))^{2}\ll m^{2}$
in the final asymptotically flat region. Between $q_{2}\simeq
m_{H}\exp(-N_{e})$ and $q_{2}\simeq \exp(-N_{e})$, we see
$\left|\beta_{q_{2}}\right|^{2}\propto q_{2}^{\ -1}$; and between
$q_{2}\simeq \exp(-N_{e})$ and $q_{2}\simeq m_{H}$, we see
$\left|\beta_{q_{2}}\right|^{2}\propto q_{2}^{\ -3}$.  In light of
these characteristics, a comparison of Eqs.~(\ref{eq:disp})
and~(\ref{eq:massdisp}) can be made with consideration to where
$(k/a(t))^{2}\gg m^{2}$ and to where $(k/a(t))^{2}\ll m^{2}$. Such
an analysis shows that in the tiny mass limit of $m_{H}\ll1$, the
dispersion spectrum reduces to the massless dispersion spectrum. }
\end{figure}

In the massless case, $P=2$, as can be seen by Eqs.~(\ref{eq:p21})
and~(\ref{eq:p22}). The height of the major peak in the graph of the
massless case in Fig.~\ref{fig:massive2pt} grows with an increasing
number of e-folds as $e^{N_{e}}$, while the widths of the peaks
narrow with an increasing number of e-folds as $e^{-N_{e}}$.  The
area under an individual peak in the massless graph therefore does
not change appreciably when changing the number of e-folds of
expansion, provided there are at least a few e-folds of inflation.
For the massive cases, we see that $P=2.93358$ when $m_{H}^{\
2}=0.1$, and that $P=2.23607$ when $m_{H}=1$.  See
Fig.~\ref{fig:pvmh}.

We wish now to approximate the dependence of the configuration space
dispersion $\left< 0 \right|\delta\phi^{2}\left| 0 \right>_{\rm
re}/H_{\rm infl}^{2}$ on the number of e-folds. This dispersion is
obtained from Eq.~(\ref{eq:2pt}) and its continuum limit. It is
proportional to the area under the curves of $\left<\
\right|\delta\phi^{2}\left|\ \right>_{q_2}/H_{\rm infl}^{2}$ in Fig.
\ref{fig:massive2pt} for the particular values of the mass shown.
The main contribution to the area under each curve comes from the
intermediate values of $q_2$ (i.e., $\exp(-N_e) < q_2 < 1$). For
this range of $q_2$, the value of $\left<\
\right|\delta\phi^{2}\left|\ \right>_{q_2}/H^{2}$ decreases as
$(q_{2}e^{N_{e}})^{2-P}$ in the massive case| or as
$(q_{2}e^{N_{e}})^{1-P}$ in the massless case| until the onset of
large-$q_{2}$ behavior at $q_{2}=1$. The values of the exponents
containing $P$ can be seen from Eqs.~(\ref{eq:disp}),
(\ref{eq:massdisp}), and~(\ref{eq:P}). (Recall that $P=2$ in the
massless case, and the range of $P$ for the massive case is shown in
Fig.~\ref{fig:pvmh}.) The value $q_2 = 1$ effectively serves as a
cut-off because of the rapid fall-off of $\left<\
\right|\delta\phi^{2}\left|\ \right>_{q_2}/H^{2}$ with increasing
$q_2$ in the large $q_2$ region.  We find that the actual height of
the major peak corresponding to the one normalized to unity in
Fig.~{\ref{fig:massive2pt} can be approximated as $J\
e^{(P-2)N_{e}}$ in the massive case, and as $J\ e^{(P-1)N_{e}}$ in
the massless case, with $J$ having the same value, $J\simeq0.01$.
Then we find that
\begin{eqnarray}
&&0.01\lesssim m_H\lesssim1.49:\nonumber\\
&&\ \ \ \ \ \frac{\left< 0 \right|\delta\phi^{2}\left| 0
\right>_{\rm re}}{H_{\rm infl}^{2}}=\frac{1}{2}J\
e^{(P-3)N_{e}}+\int_{e^{-N_{e}}}^{1}dq_{2}\ J\
q_{2}^{2-P},\nonumber\\[2mm]
&&m_H=0:\nonumber\\
&&\ \ \ \ \ \frac{\left< 0 \right|\delta\phi^{2}\left| 0
\right>_{\rm re}}{H_{\rm
infl}^{2}}=\frac{1}{2}J+\int_{e^{-N_{e}}}^{1}dq_{2}\ J\ q_{2}^{1-P}.
\end{eqnarray}
The resulting configuration space dispersion is given in
\nobreak{TABLE~\ref{tab:area}}.
\begin{table}
\caption{Configuration Space Dispersion} \label{tab:area}
\begin{tabular}
{ c | c |} & $\left< 0 \right|\delta\phi^{2}\left| 0 \right>_{\rm
re}/H_{\rm infl}^{2}$\\  \hline
$0.01\lesssim m_H\lesssim1.49$ & $\left(\frac{1}{3-P}+\frac{1-P}{6-2P}e^{(P-3)N_{e}}\right)J$\\
\hline $m_H=0$ & $\left(\frac{1}{2}+N_{e}\right)J$\\
\hline
\end{tabular}
\end{table}
From this, one can deduce that
the small mass limit, for which $P\rightarrow3$, gives the
same result for $\left|\delta\phi\right|$, as one has in
the massless case, namely
\begin{equation}
\left|\delta\phi\right| \equiv \sqrt{\left< 0
\right|\delta\phi^{2}\left| 0 \right>_{\rm re}} \simeq\frac{H_{\rm
infl}}{10}\sqrt{N_{e}+\frac{1}{2}}.
\end{equation}
From the behavior discussed in the caption to
Fig.~\ref{fig:tinymass}, one can show that this continuity
also holds for each mode separately.

\section{Spectral Index}
\label{sec:spec}

We say that a given mode $k$ of the perturbation field $\delta\phi$
is crossing the Hubble radius at the time when $k/(a(t)H(t))=1$. For
larger values of $k/(a(t)H(t))$ (shorter wavelengths) the mode is
said to be inside the Hubble radius, and for smaller values it is
said to be outside the Hubble radius. Modes in the
intermediate-$q_{2}$ range, as defined after Eq.~(\ref{q2}), exit
during inflation to eventually re-enter the Hubble radius at some
time after inflation has ended. Using our composite scale factor, we
note that after a few e-folds of inflation, the quantum
perturbations that are exiting the Hubble radius are found
numerically to satisfy|
\begin{eqnarray}
\left|\psi_{k}\right|^{2}=\frac{H_{\rm infl}^{2}}{k^{3}D(m_{H})},
\label{eq:scaleinv}
\end{eqnarray}
where it will be recalled that $\left|\delta\phi_{k}^{\rm
(un)}\right|^{2} = (2\pi)^{-3}\left|\psi_{k}\right|^{2}$, as given
by Eqs.~(\ref{eq:phi}) and~(\ref{eq:wave}). The variable $D(m_{H})$
is a constant of order 1 that we have evaluated numerically to be|
\begin{eqnarray}
D(m_{H}=0)&=&1.00,
\nonumber\\
D(m_{H}=\sqrt{0.1})&=&1.04,
\nonumber\\
D(m_{H}=1)&=&1.45.
\label{eq:Dvalues}
\end{eqnarray}
One finds approximately that
$D(m_{H})\simeq(1+\frac{1}{5}m_{H}^{\ 2})^{2}$.
A few Hubble times after the mode has exited the
Hubble horizon, the value of $\left|\psi_{k}\right|^{2}$
approaches a constant value of
$1/2$ that given in Eq.~(\ref{eq:scaleinv}) when
$m_{H}=0$.  In the case when $m_H\ll 1$,
the value of $\left|\psi_{k}\right|^{2}$ approaches about
$1/2$ the value in Eq.~(\ref{eq:scaleinv}) after a few
Hubble times, but then decreases very slowly over
many Hubble times.

The scalar spectral index $n_s$ is defined by
\begin{equation}
n_{s}=1+\frac{d\ln\mathcal{P}_{\delta\phi}}{d\ln k}.
\label{eq:spectral}
\end{equation}
At the time of exit from the horizon, or at several Hubble times
after exit, we see that the spectrum defined by
Eq.~(\ref{eq:defp}) satisfies the proportionality
\begin{equation}
\mathcal{P}_{\delta\phi}\propto k^{3}\left|\delta\phi_{k}^{\rm
(re)}\right|^{2}, \label{eq:spectrum}
\end{equation}
with $\left|\delta\phi_{k}^{\rm (re)}\right|^{2}\propto k^{-3}$.
Therefore the spectral index $n_s$ is 1 for constant $H$ as measured
at any given number of Hubble times after horizon exit during
inflation.  This is true for the range of values of $m_H$ considered
in Eq.~(\ref{eq:Dvalues}).

The modes that exit the Hubble radius at early times
before, or shortly after,
the scale factor begins growing exponentially are not described
by Eq.~(\ref{eq:scaleinv}).  These modes, which
are primarily in the small-$q_{2}$ region, are not scale-invariant.
The gravitational perturbations induced by these modes
could reenter the Hubble radius during the
matter-dominated or dark-energy dominated stages of
expansion, and would appear as long wavelength modes
having a spectrum that is not scale invariant.
If the total number of e-folds of
inflation is sufficiently small, it would be possible to observe a
breaking of the scale-invariance of the large-scale structure
of the universe at sufficiently large scales.
See Figs.~\ref{fig:b2vq2},
\ref{fig:massive}, and~\ref{fig:domterm}, in which the breaking
of scale-invariance is seen in the small-$q_{2}$ region.
The curves of $\log( |\beta_{q_2}|^2)$ have a slope of $-2$ in
the scale-invariant region of intermediate values of $q_2$ and
a slope of $0$ in the region of small $q_2$ values.

Because the small-$q_{2}$
modes of large enough wavelength exit the Hubble radius before
evolving away from the early-time conditions specified by
Eq.~(\ref{eq:minkowski}), we would expect a massless inflaton to
generate a spectral index of $n_{s}=3$ in the small-$q_{2}$ region.
(This follows from the fact that the square of the amplitude in
Eq.~(\ref{eq:minkowski}) goes as $k^{-1}$ for 0 mass.)
Similarly, we would expect a massive inflaton to
generate a spectral index
of $n_{s}=4$ in the small-$q_{2}$ region, if $m_H$ satisfies
$q_{2}\ll\exp(-N_{e})\,m_{H}$ in that region.
If scale-invariance continued
indefinitely for large wavelength modes, the dispersion would be
infrared divergent, so we expect an end to scale-invariance to occur
at very large length scales for any reasonable spectrum of
inflaton perturbations. Let us examine this breaking of scale
invariance for the $a(t)$ we have been considering.
The modes responsible for
galaxy-size structures today, left the Hubble radius approximately
45 e-folds before the end of inflation \cite[pp.~285]{KT}, so if
$N_{e}$ were not too much larger than 45, we would expect it to be
possible to observe the breaking of scale invariance at sufficiently
large scales within our observable universe.

It is of interest to understand the behavior of the coefficients
$E(k)$ and $F(k)$ in the expression, given in
Eq.~(\ref{eq:massinfleigen}), of the
mode functions as a superposition of Hankel functions.  These
coefficients determine the quantum state of the inflaton
perturbation field $\delta\phi$.  Because the modes
evolve separately in this model, the state vectors can be
considered for each mode separately.  For the modes that
exhibit a scale-invariant spectrum, we will see that the
quantum state is the Bunch-Davies vacuum during inflation
and that it evolves naturally from the initial Minkowski
vacuum that we choose in the early time asymptotically
flat segment of the scale factor. We find numerically that
$\left|E(k)\right|\sim1$ and $F(k)\sim0$ for modes of
intermediate-$q_{2}$, which are the modes that exit during the
exponential expansion of our composite scale factor. Some sample
values of $E(k)$ and $F(k)$ for our composite scale factor are given
in TABLE~\ref{tab:hankcoef}, where the values of $q_{2}$ listed
assume a value of $N_{e}=60$.  The first row of data is in the
small-$q_{2}$ region. The second
row of data is roughly at the interface between small-
and intermediate-$q_{2}$, and the fifth row of data
is at the interface between intermediate- and large-$q_{2}$.

Thus, for intermediate-$q_2$ modes (the ones that exit the horizon
after inflation has started), and large-$q_2$ modes (the ones that
never exit the horizon) the value of $|E(k)|$ is essentially 1 and
the value of $|F(k)|$ is essentially 0. This corresponds to the
Bunch-Davies vacuum \cite{bunch-davies} in de Sitter spacetime.
Since $E(k)$ and $F(k)$ are constant in time, this fact must be a
consequence of our choice of state in the initial asymptotically
flat spacetime at very early times. We took that state to be the
Minkowski vacuum, having no particles present at early times. For
sufficiently high frequency modes, there are no particles created by
the initial expansion prior to the time $t_1$ when we join it to the
inflationary segment (i.e., prior to the time when inflation begins
in our model).  The rate of expansion at this joining is equal to
$H_{\rm infl}$, so one would expect few
particles to be present at time $t_1$ in modes for which
the momentum satisfies
$k/a(t_1) \gtrsim H_{\rm infl}$.  This condition implies
that the number of created particles present at the joining
to inflation is negligible at values
of $q_2$ larger than those in the small-$q_2$ region.
Thus, it is reasonable\footnote{For a discussion of the
Bunch-Davies vacuum in connection with the energy-momentum
tensor see \cite{Andersonetal}. }
that the
initial Minkowski vacuum goes over into what is essentially the
Bunch-Davies vacuum for values of $k$ that satisfy this condition
at the time inflation starts in our model.

\begin{table}
\caption{Sample Hankel Coefficients for $N_e=60$}
\label{tab:hankcoef}
\begin{tabular}
{| c | c | c|} $q_{2}$ & $E(k)$ & $F(k)$\\ \hline $10^{-30}$ &
$-4378.26-0.5i$ & $-4378.26-0.5i$\\
\hline
$10^{-26}$ & $-1.0915-0.017581i$ & $-0.18204-0.39819i$\\
\hline
$10^{-23}$ & $0.99924-0.038888i$ & $-1.68347\times10^{-5}+4.37502i\times10^{-4}$\\
\hline
$10^{-13}$ & $-0.52204+0.85292i$ & $3.73431\times10^{-14}-2.28562i\times10^{-14}$\\
\hline
$1$ & $-0.9082-0.41854i$ & $2.09333\times10^{-27}-2.78876i\times10^{-27}$\\
\hline
$10^{6}$ & $0.16978-0.98548i$ & $6.61571\times10^{-45}-1.13990i\times10^{-45}$\\
\hline
\end{tabular}
\end{table}

The large-$q_2$ modes do not exit the Hubble horizon before
inflation ends with the joining to the late-time asymptotically
Minkowskian region at $t_2$ in our model.  Hence, for those modes
one still has at late times that $k/a(t_2) > H_{\rm infl}$, so in
those modes there are no created particles of the field
$\delta\phi$, i.e., no perturbations of the inflaton field are
created. This behavior can be seen in Figs.~\ref{fig:b2vq2},
\ref{fig:massive}, and~\ref{fig:domterm}, where a rapid fall off in
$|\beta_k|^2$ is seen for the large-$q_2$ region. By a similar
argument to that in the previous paragraph, we would expect the
number of particles of $\delta\phi$ to be negligible in the final
Minkowski region for these modes $k$ that did not exit the Hubble
horizon during inflation. For these modes, the Bunch-Davies vacuum
goes over into the Minkowski vacuum at late times.

On the other hand, for intermediate-$q_2$ modes $a(t_2)/a(t_1)$ is
sufficiently large that $k/(a(t_2)H_{\rm infl})\lesssim 1$.
Therefore, we would expect the intermediate-$q_2$ modes to contain
created particles of the $\delta\phi$ field in the final Minkowski
space. This can be confirmed from Figs.~\ref{fig:b2vq2},
\ref{fig:massive}, and~\ref{fig:domterm}.

For these intermediate values of $q_2$, the value of the argument
$z$ of the Hankel functions is small in the late stage of inflation.
Therefore, one can use the asymptotic form,
\begin{equation}
\left|H_{v}^{(1)}(z)\right|^{2}\simeq\left|H_{v}^{(2)}(z)\right|^{2}\simeq\left(\frac{\Gamma(v)}{\pi}\right)^{2}\left(\frac{1}{2}z\right)^{-2v},
\label{eq:smallargHankel}
\end{equation}
in Eq.~(\ref{eq:massinfleigen}) to find that
$\left|\psi_{k}\right|^{2}\simeq
a^{-3}\left|H_{v}^{(1)}(z)\right|^{2}\propto a^{-3}z^{-2v}$, where
$z=k/(a(t)H_{\rm infl})$ and $v=\sqrt{(9/4)-m_{H}^{\ 2}}$.
From Eq.~(\ref{eq:smallargHankel}), we see that
$\left|\psi_{k}\right|^{2}$ is constant when $m_{H}=0$ and
decreases with time as $a(t)^{(\sqrt{9 - 4m_{H}^{\ 2} } - 3) }$
for $ 3/2 > m_{H}^{\ 2} >0$.

Using
this small argument approximation with Eq.~(\ref{eq:spectrum})
leads to
\begin{equation}
\frac{d\ln\mathcal{P}_{\delta\phi}}{d\ln k}=3-\sqrt{9-4m_{H}^{\ 2}},
\end{equation}
Then Eq.~(\ref{eq:spectral})
gives a spectral index of
\begin{equation}
n_{s}=4-\sqrt{9-4m_{H}^{\ 2}}.
\label{eq:blue}
\end{equation}

When $m_{H} =0$, this gives the scale-invariant spectrum of
inflaton fluctuations corresponding to $n_{s}=1$. The slow-roll
approximation will reduce the value of $n_{s}$ because
$H_{\rm infl}$
decreases with time. This easily could be incorporated into our
model, but will not be pursued further here.

For $3/2 > m_{H}^{\ 2} >0$, Eq.~(\ref{eq:blue}) would
seem to give $n_{s}>1$, but this is because in obtaining
that equation we have used the same
given $t$ for all modes, which means that $z$ is proportional
to $k$ for each mode.
However, the perturbation spectrum of
$\delta\phi$ is believed to induce the spectrum of
scalar perturbations of the metric when the modes are at
a given number of wavelengths outside the inflationary
Hubble horizon \cite{LL, Dodelson}. This corresponds to
the same value of
$z$, not $t$, for each mode. When that is taken into account,
the quantity $d\ln\mathcal{P}/{\ln k} = 0$, thus giving the
scale-invariant value $n_{s}=1$ for any value of $m_{H}$.

Because the gravitational perturbations are effectively
massless, if one were to include them and
propagate them to late-times using the scale-factor $a(t)$
in our model, one would find the same scale-invariant
spectrum as we obtained for an inflaton perturbation field
with $m_{H} \ll 1$.

The graph of the inflaton spectra at late times
for intermediate $q_2$ in Fig.~{\ref{fig:lnMassDispSpec} }
has the spectral index of Eq.~(\ref{eq:blue}) because
the joining to the asymptotically flat late time region occurs
at a given time $t_2$.  For massless inflatons, as for massless
metric perturbations,  the fact that inflation ends approximately
at a given time does not affect the scale-invariance of the
spectrum.\footnote{The scale-invariance of the initial spectrum of
scalar perturbations of the metric when inflation ends, thus
places an upper limit on the graviton mass, at least in principle.}

\section{Density Perturbations}
\label{sec:dens}

In this section, we assume as usual that the inflaton perturbation
field $\left|\delta\phi_{k}^{\rm (re)}\right|$ sets up scalar
perturbations of the metric shortly after their wavelengths exit the
Hubble horizon. We will consider two scenarios.

In the first scenario, the
modes of the quantized inflaton field directly give rise to the
scalar metric perturbations, without taking into account any
regularization or renormalization of the ultraviolet divergences of
the dispersion (or variance) of the quantized inflaton perturbation
field, beyond the Minkowsi vacuum energy subtraction.  If there
are no further subtractions during the
time when the scalar metric perturbations are induced, then the
dispersion of the quantized inflaton field has an ultraviolet
divergence. It is only in the late-time
Minkowski space that this subtraction would give a finite dispersion
for the quantized inflaton field.\footnote{This can be seen, for
example, by looking at the terms in the adiabatic subtraction that
vanish when $\dot a$ and $\ddot a$ are $0$ -- those terms are
divergent when integrated over all modes.}  Thus UV divergences
are ignored in this first scenario,
presumably because the modes relevant
to observable scales today are assumed to be unaffected by
renormalization of the ultraviolet divergences in the dispersion.
However, it is not at all obvious that this is correct. This will
become evident when we consider the second scenario.

In the second scenario, the modes of
the quantized inflaton field will again induce scalar metric perturbations,  but we will regularize or
renormalize the dispersion of the ultraviolet divergences
of the quantized inflaton field, including the renormalization
terms that are required for the dispersion to remain finite
during the expansion of the universe, and not just the terms
required in the late time Minkowski limit.  We assume that at a
time shortly after exit from the Hubble horizon, the scalar
perturbations of the metric are induced and may be treated
as essentially classical after that time \cite{Kiefer, Agullo2}.
It is most straightforward to use adiabatic regularization
\cite{Parker5} to find the dispersion because it gives the result
explicitly for each mode of the
quantized inflaton perturbations.  It should not make a significant
difference to our result if one were to use, for example, Hadamard
point-splitting regularization to obtain the
dispersion \cite{BiLuPiJuAn}.
One way to understand the rational for adiabatic
regularization is analogous to that originally given
in \cite{Parker2, Parker3} in which
a measuring instrument was considered.  In the present case,
one may suppose that the classical gravitational metric
acts like a classical measuring instrument that cannot support
the ultraviolet divergences that arise from the short-wavelength modes  of the quantized inflaton field. The natural subtraction procedure that
arises from this assumption is embodied by adiabatic regularization.

We now evaluate the expected dispersion of the
scalar metric perturbations by both methods for the simple inflaton
potential quadratic in the inflaton field. We work with the case when
$H_{\rm infl}$ is constant.  As in \cite{Parker1, Agullo2},
we find that when regularization is taken into account, the relevant
metric perturbations induced by a given value of $H_{\rm infl}$ are
considerably smaller than they would be if no regularization
of the quantized inflaton field were
taken into account. (By ``relevant modes,'' we mean those responsible
for temperature variations in the CMB and large scale structure that
we observe today.) Put another way, when regularization is taken
into account, the value of $H_{\rm infl}$ responsible for the these
observed variations must be larger than it would be without
regularization.

The basic reason for the smaller dispersion found in the
second scenario for the scalar metric perturbations
is that the metric perturbations are induced,
and acquire classical properties, within about a Hubble time
after the inflaton perturbations have left the horizon.
The dispersion of the
classical scalar metric perturbations depends on the value
of the dispersion of the inflaton field at the time when the
perturbations are induced. Because that time is soon after
the inflaton perturbations have exited the Hubble horizon, the
adiabatic subtractions are quite significant. After that time
the induced scalar metric perturbations are treated as classical,
requiring no further subtractions.

If the scalar metric perturbations were induced only
at the end of the inflationary era,  then the
relevant modes of the inflaton perturbation field would
have left
the Horizon a large number of Hubble times earlier and the
adiabatic subtractions would be small.  In that case, the adiabatic subtractions would give only a small difference from the
first scenario. However, in the second scenario we are
assuming the metric perturbations are induced soon after
horizon exit.

In the case when $m_H{}^2 \ll 1$, we find numerically that the
duration and manner in which inflation ends does not have much
affect on the value of $\left|\delta\phi_{k}^{\rm (un)}\right|^{2}$,
after a given mode in the relevant range of wavelengths has exited
the Hubble radius. Thus, the value of $\left|\delta\phi_{k}^{\rm
(un)}\right|^{2}$ at late times is approximately equal to the value
it has soon after leaving the Hubble horizon. At late times,
Eqs.~(\ref{eq:wave}),~(\ref{eq:alphabeta}),
and~(\ref{eq:Bogoliubov}) show that the expectation value is
\begin{eqnarray}
{\left|\delta\phi_{k}^{\rm
(un)}\right|^{2}}&&=\frac{1}{2L^{3}a_{2f}^{\ \
3}\omega_{2f}}\left(\left|\alpha_{k}\right|^{2}+\left|\beta_{k}\right|^{2}\right)
\nonumber\\
&&=\frac{1}{2L^{3}a_{2f}^{\ \
3}\omega_{2f}}\left(1+2\left|\beta_{k}\right|^{2}\right).
\label{eq:timeavgphi}
\end{eqnarray}
The value of $\left|\delta\phi_{k}^{\rm (un)}\right|^{2}$ obtained
from Eqs.~(\ref{eq:phi}), (\ref{eq:wave}), and~(\ref{eq:scaleinv}),
however, is un-renormalized.  In the late time Minkowski spacetime,
the renormalization consists of subtracting the vacuum contribution
corresponding to the late time Minkowski spacetime for which the
scale factor has the value $a_{2f}$. This is equivalent to replacing
$(1+2\left|\beta_{q_{2}}\right|^{2})$ by
$2\left|\beta_{q_{2}}\right|^{2}$ in Eq.~(\ref{eq:timeavgphi}), as was
done to obtain Eq.~(\ref{eq:MinkSubtraction}).  Thus, at late times
\begin{eqnarray}
{\left|\delta\phi_{k}^{\rm
(re)}\right|^{2}}=\frac{1}{L^{3}a_{2f}^{\ \ 3}\omega_{2f}}
\left|\beta_{k}\right|^{2}.
\label{eq:timeavgphiRenorm}
\end{eqnarray}

For intermediate values of $q_2$ in the massless case,
it follows from Eq.~(\ref{eq:p22}) that at late times
\begin{equation}
\left|\beta_{q_{2}}\right|^{2}\simeq\frac{1}{4}q_{2}{}^{-2}.
\label{eq:betaq2squared}
\end{equation}
We write the renormalized spectrum of $\delta{\phi}(x)$  in
terms of the modes of the renormalized expectation value
of the dispersion (or variance),
$\langle0\left|(\delta{\phi}(x) ){}^2\right|0\rangle_{\rm re}$,
which are defined in Eq.~(\ref{eq:defRenormModes}).
This gives the spectrum as
\begin{eqnarray}
\mathcal{P}_{\delta\phi^{(re)}} &\equiv&
\left(\frac{L}{2\pi}\right)^{3}4\pi\,k^{3}
\langle0 |(\delta{\hat \phi}_{\vec k} ){}^2 |0\rangle_{\rm re}\nonumber\\
&=&
\left(\frac{L}{2\pi}\right)^{3}4\pi\,k^{3}
\left| \delta\phi^{\rm (re)}_{k}\right|^{2}.
\label{eq:spectrumDeltaphiRe}
\end{eqnarray}
From Eqs.~(\ref{eq:timeavgphiRenorm}) and
(\ref{eq:betaq2squared}), we obtain, in the
continuum limit, the late-time
renormalized inflaton perturbation spectrum
\begin{equation}
\mathcal{P}_{\delta\phi^{(re)}}=\frac{H_{\rm infl}^{2}}{8\pi^{2}}.
\end{equation}
From our discussion of Fig.~\ref{fig:tinymass}, this result also
holds in the massive case when $m_{H}{}^{2}\ll1$.  Here, in the
range where $m^{2}\gg k^{2}/a_{2f}{}^{2}$, we have
$\left|\beta_{q_{2}}\right|^{2}\propto q_{2}{}^{-3}$.  As can be
seen in Fig.~\ref{fig:tinymass}, this region is continuous with the
massless case given by Eq.~(\ref{eq:betaq2squared}) for
$k^{2}/a_{2f}{}^{2}\gg m^{2}$.  This continuity tells us that the
region of $\left|\beta_{q_{2}}\right|^{2}$ proportional to
$q_{2}{}^{-3}$ must follow
\begin{equation}
\left|\beta_{q_{2}}\right|^{2}\simeq\frac{m_{H}}{4}q_{2}{}^{-3},
\end{equation}
and thus
\begin{equation}
\mathcal{P}_{\delta\phi^{(re)}}=\frac{H_{\rm infl}^{2}}{8\pi^{2}},
\label{eq:deltaphiren}
\end{equation}
for both the massless case and the case where $m_{H}{}^{2}\ll1$.
This is the result for the spectrum in the first scenario, in which
only the late-time Minkowski vacuum subtractions are considered
and the renormalization of the inflaton dispersion during the
expansion of the universe is ignored.

However, by late-times it is usually assumed that the inflaton
perturbations have already induced the scalar metric perturbations
that will give rise to the initial scale-invariant plasma oscillations
in the reheated universe.  Because the universe is still expanding
rapidly at the time that the relevant scalar metric perturbations are
induced, one should consider the effect of renormalization of the
inflaton perturbations in
the rapidly expanding universe on the amplitude of the
metric perturbations that they induce.  Following
\cite{Parker1, Agullo, Agullo2}, we take the scalar
metric perturbations
to be classical in nature shortly after they
are induced.  Therefore, in our second scenario,
it is the renormalized spectrum
of the quantized inflaton perturbation field shortly after
their wavelengths exit the Hubble horizon during inflation
that we use to determine the magnitude of the spectrum of
the induced scalar metric perturbations. The latter propagate as
classical quantities during the long remaining duration
of inflation, and therefore their amplitudes remain almost
constant up to the time that they induce acoustic
plasma oscillations after reheating.  The adiabatic
subtractions coming from renormalization are significant
at the time that the scalar perturbations of the metric
are induced, and hence have an important influence on
the magnitude of the metric perturbations and consequently
on the CMB anisotropies and the large scale structure observed
today. Slow-roll inflation will
alter the scale-invariant spectrum, as considered
with the effect of renormalization in \cite{Agullo2}.
The effect of renormalization is found there to be
important for interpreting the existing measurements
of the CMB anisotropies.

Next, we consider the effect of
renormalization on the amplitude of the spectrum
of scalar metric perturbations induced
by the quantized inflaton field during inflation. We give
the results for both scenarios; the first, in which only the
Minkowski subtractions are made and the late-time dispersion
in the final Minkowski space is used; and the second,
in which the renormalization is carried out
near the time when the wavelength of the mode
first exceeds the Hubble radius during inflation.

The curvature perturbations (or scalar
perturbations) of the metric
\cite{LL} are defined by
\begin{equation}
\mathcal{R}_{k}=-\frac{H}{\dot{\phi}}
\left | \delta\phi_{k}^{\rm (re)} \right |,
\label{eq:CurvaturePert}
\end{equation}
where $\dot{\phi}$ represents $\dot{\phi}{}^{(0)}$ in the
present section.
In \cite{LL}, the possible effects of renormalization by adiabatic
regularization or point-splitting were not considered, and the
unrenormalized value $\left | \delta\phi_{k}^{\rm (un)} \right |$
was used.  However, here we will use the renormalized value, as it
is clear that at late times one must subtract the Minkowski vacuum
contribution. This subtraction is already included in adiabatic
regularization (and in point-splitting) when applied to Minkowski
space. However, when $\dot a$ or $\ddot a$ are nonzero, the simple
Minkowski space subtraction is not enough to yield a finite
value for the dispersion of the inflaton fluctuation field in the
expanding universe.  As noted
in \cite{Parker1,  Agullo, Agullo2}, the effect
of renormalization in the expanding universe may have a
significant effect on the magnitude of the inflaton
dispersion spectrum.  As discussed in the previous paragraph,
we take the relevant time at which the magnitude of the
inflaton perturbation field induces scalar metric perturbations to
be shortly after horizon crossing.  Therefore,
in the second scenario we use the
horizon crossing time as a characteristic time at which to
evaluate the renormalized spectrum in
Eq.~(\ref{eq:CurvaturePert}).  One will obtain a similar result
as long as the characteristic time is taken within a few
Hubble times after horizon crossing.

The spectrum,
$\mathcal{P}_{\mathcal{R}}=(L/2\pi)^3 4\pi k^3 {\mathcal{R}_k}^2$,
of scalar metric perturbations in the continuum limit
now takes the form
\begin{equation}
\mathcal{P}_{\mathcal{R}}=\frac{H^{2}}{\dot{\phi}^{2}}\mathcal{P}_{\delta\phi^{(re)}}.
\label{eq:PsubR}
\end{equation}
The difference between the two scenarios boils down to the
time at which the renormalized dispersion is calculated.

So far, our parameterized scale factor has not been linked to any
particular potential or model of inflation.  In what comes next, we
choose a simple potential that was found to be in good agreement
with the 3-year WMAP data \cite{3WMAP}. For our example, we use the
quadratic chaotic-inflation potential \cite{Linde}
\begin{equation}
V=\frac{1}{2}m^{2}\phi^{2}.
\label{eq:pot}
\end{equation}
The two slow roll conditions are \cite{SR}
\begin{equation}
H^{2}\simeq\frac{8\pi G}{3}V,
\end{equation}
and
\begin{equation}
\dot{\phi}\simeq-\frac{dV/d\phi}{3H}.
\end{equation}
We combine these two slow roll equations with the
present quadratic potential
to find
\begin{equation}
\dot{\phi}\simeq-m\sqrt{\frac{2}{3}}\ \frac{1}{\sqrt{8\pi G}}.
\end{equation}
Then with the quadratic potential, in the first scenario, we find
from Eqs.~(\ref{eq:deltaphiren}) and (\ref{eq:PsubR}) with only the
late-time Minkowski subtraction, that
\begin{equation}
\mathcal{P}_\mathcal{R}=\left[\frac{3H_{\rm infl}^{2}}
{2m^{2}\ \left({1/\sqrt{8\pi G}}\right)^{2}}
\right] \frac{H_{\rm infl}^{2}}{8\pi^{2}},
\end{equation}
which can be written as
\begin{equation}
\mathcal{P}_\mathcal{R}=\frac{3}{16\pi^{2}\,m_{H}{}^{2}}\left(\frac{H_{\rm
infl}}{2.436\times10^{18}GeV}\right)^{2}.
\end{equation}

In the second scenario, we take account of the further
subtractions necessary to renormalize the inflaton dispersion
when the time-derivatives of $a(t)$ cannot be neglected and
obtain for the inflaton dispersion spectrum.
We use the result given in \cite{Parker1} for the
renormalized spectrum, $\mathcal{P}_{\delta\phi^{(re)}}$.
As noted before, this should be evaluated shortly after the
mode has exited the Hubble radius.  For simplicity, we use
the time of exit to characterize the order of magnitude one will
obtain. This gives
\begin{eqnarray}
\mathcal{P}_{\delta\phi^{(re)}}&\approx&\frac{H_{\rm infl}^2}{32\,\pi^{2}}
\bigg(4\pi\left|H_{\sqrt{\frac{9}{4}-m_{H}{}^{2}}}^{(1)}\left(1\right)\right|^{2}\nonumber\\
&&-\frac{8m_{H}{}^{6}+33m_{H}{}^{4}+46m_{H}{}^{2}+16}{\left(m_{H}{}^{2}+1\right)^{7/2}}\bigg),\
\ \ \ \ \ \ \ \
\label{eq:inflatonSpectrumAdiabaticReg}
\end{eqnarray}
from which one obtains,
\begin{eqnarray}
\mathcal{P}_{\mathcal{R}}&\approx&\frac{3}{64\,\pi^{2}\,m_{H}{}^{2}}\left(\frac{H_{\rm
infl}}{2.436\times10^{18}GeV}\right)^{2}\nonumber\\
&&\times\bigg(4\pi\left|H_{\sqrt{\frac{9}{4}-m_{H}{}^{2}}}^{(1)}\left(1\right)\right|^{2}\nonumber\\
&&-\frac{8m_{H}{}^{6}+33m_{H}{}^{4}+46m_{H}{}^{2}+16}{\left(m_{H}{}^{2}+1\right)^{7/2}}\bigg).\
\ \ \ \ \ \ \ \
\end{eqnarray}

After inflation has ended, at the time when
once again $k/(a(t)H(t))=1$, the
curvature perturbations reenter the Hubble sphere.
At that reentry time, their amplitudes are related to
the amplitudes, $\delta\rho_{k}$, of the density
perturbations by relations of the form \cite{LL},
\begin{equation}
\frac{\delta\rho_{k}}{\rho}\equiv\delta_{k}=
\chi\,\mathcal{R}_{k},
\end{equation}
where $\chi$ is $2/5$ if reentry occurs during the matter-dominated
stage, and is $4/9$ if reentry occurs during the radiation-dominated
stage. We use the value $2/5$ in the rest of this section. Then, in
the first scenario, the specturm of density perturbations created
by the scalar metric perturbations at the time of reentry is
\begin{eqnarray}
\mathcal{P}_{\delta}&=&\left(\frac{2}{5}\right)^{2}\mathcal{P}_{\mathcal{R}}\nonumber\\
&=&\frac{3}{100\,\pi^{2}\,m_{H}{}^{2}}\left(\frac{H_{\rm
infl}}{2.436\times10^{18}GeV}\right)^{2}.
\label{eq:firstscen}
\end{eqnarray}
In the second scenario, assuming that the magnitude of
the scalar metric perturbations does not change much from the
value it has at the time that the metric perturbations are induced
to the time of reentry, we have
\begin{eqnarray}
\mathcal{P}_{\delta}&\approx&\frac{3}{400\,\pi^{2}\,m_{H}{}^{2}}\left(\frac{H_{\rm
infl}}{2.436\times10^{18}GeV}\right)^{2}\nonumber\\
&&\times\bigg(4\pi\left|H_{\sqrt{\frac{9}{4}-m_{H}{}^{2}}}^{(1)}\left(1\right)\right|^{2}\nonumber\\
&&-\frac{8m_{H}{}^{6}+33m_{H}{}^{4}+46m_{H}{}^{2}+16}{\left(m_{H}{}^{2}+1\right)^{7/2}}\bigg).\
\ \ \ \ \ \ \ \
\label{eq:secondscen}
\end{eqnarray}

See TABLE~\ref{tab:fluc} for sample values of the density contrast,
$\mathcal{P}_{\delta}$, given by Eq.~(\ref{eq:firstscen}) in the
first scenario, and see TABLE~\ref{tab:fluc2} for sample
values of the
density contrast given by Eq.~(\ref{eq:secondscen})
in the second scenario.
In both cases, the potential is the quadratic one
of Eq.~(\ref{eq:pot}).
The quantity $\mathcal{P}_{\delta}$ in our table is related to
the measured quantity denoted by $\Delta^2_{\mathcal{R}}$
in the WMAP five-year results \cite[see their Table 2]{Dunkley}.
Their measurements give for the most likely value,
$\Delta^2_{\mathcal{R}} = 2.41\times 10^{-9}$.
In our present paper, $\Delta^2_{\mathcal{R}}$ is denoted
by $\mathcal{P}_{\mathcal{R}}$, and from the first line of
Eq.~(87), we obtain its observed value:
$\mathcal{P}_{\delta} = (2/5)^2 \mathcal{P}_{\mathcal{R}}
\approx 3.86 \times 10^{-10}$.
We have chosen $H$ to give values of $\mathcal{P}_{\delta}$ in the first row of each table that are near the observed value of
$\mathcal{P}_{\delta}$.

A comparison of TABLE~\ref{tab:fluc2} with
TABLE~\ref{tab:fluc} reveals that adiabatic regularization leads to
a constant non-zero value of $P_{\delta}$ for small values of
$m_{H}$.  Thus, for the quadratic potential, all the values
of $m_H$
shown for adiabatic regularization in TABLE~\ref{tab:fluc2} give a
value of $P_{\delta}$ that agrees with the observed value
when $H\approx 10^{15}$ GeV. In TABLE~\ref{tab:fluc}, we see
that $P_{\delta}$ grows to as $m_H$ decreases. For the quadratic
potential it approaches infinity in the limit as $m_H \rightarrow 0$
without adiabatic regularization, but approaches a constant non-zero value when the full adiabatic regularization is used during the
expansion. This can be confirmed by expanding the right-hand-side
of Eq.~(\ref{eq:secondscen}) in powers of $m_H$ and taking the
limit as $m_H\rightarrow 0$. The fact that the value of
$P_{\delta}$ is almost constant in TABLE~\ref{tab:fluc2} for
all the values of $m_H \le 0.25$ in each column, implies
that one could predict the value of $H_{\rm infl}$ for this,
and perhaps also for other inflaton potentials, when adiabatic
regularization is taken into account.

\begin{table}
\caption{Late-Time Method: $\mathcal{P}_{\delta}$ for
$V=\frac{1}{2}m^{2}\phi^{2}$} \label{tab:fluc}
\begin{tabular}{ l | c | c | c | c |} & $H=10^{10}\textrm{GeV}$ & $H=10^{11}\textrm{GeV}$ & $H=10^{12}\textrm{GeV}$\\  \hline
$m_H=0.0001$ & $5.12\times 10^{-12}$ & $5.12\times
10^{-10}$ & $5.12\times 10^{-8}$\\
\hline $m_H=0.01$ & $5.12\times 10^{-16}$ & $5.12\times 10^{-14}$
& $5.12\times 10^{-12}$\\
\hline $m_H=0.1$ & $5.12\times 10^{-18}$ & $5.12\times 10^{-16}$ &
$5.12\times 10^{-14}$\\
\hline $m_H=0.25$ & $8.20\times 10^{-19}$ & $8.20\times 10^{-17}$ &
$8.20\times 10^{-15}$\\
 \hline
\end{tabular}
\end{table}

\begin{table}
\caption{Adiabatic Regularization: $\mathcal{P}_{\delta}$
for $V=\frac{1}{2}m^{2}\phi^{2}$} \label{tab:fluc2}
\begin{tabular}{ l | c | c | c | c |} & $H=10^{14}\textrm{GeV}$ & $H=10^{15}\textrm{GeV}$ & $H=10^{16}\textrm{GeV}$\\  \hline
$m_H=0.0001$ & $4.60\times 10^{-12}$ & $4.60\times
10^{-10}$ & $4.60\times 10^{-8}$\\
\hline $m_H=0.01$ & $4.60\times 10^{-12}$ & $4.60\times 10^{-10}$
& $4.60\times 10^{-8}$\\
\hline $m_H=0.1$ & $4.64\times 10^{-12}$ & $4.64\times 10^{-10}$ &
$4.64\times 10^{-8}$\\
\hline $m_H=0.25$ & $4.79\times 10^{-12}$ & $4.79\times 10^{-10}$ &
$4.79\times 10^{-8}$\\
\hline
\end{tabular}
\end{table}

\section{Reheating}

\label{sec:reheat}

When we
maintain continuity of $a(t)$, $\dot{a}(t)$, and $\ddot{a}(t)$, the
particle number in a given mode is proportional to $q_{2}^{\ -6}$
for large enough values of $q_{2}$.
[See Eq.~(\ref{eq:rapidFall}).]

When the end of inflation is gradual, we found (as seen in
Fig.~\ref{fig:domterm}) that beyond $q_{2}\simeq1$, the quantity
$\left|\beta_{q_{2}}\right|^{2}$ falls off as $q_{2}{}^{-6}$.
For small $m_H$, using the
behavior of $\left|\beta_{q_{2}}\right|^{2}$
summarized in Eqs.~(\ref{eq:p21}), (\ref{eq:p22}),
and (\ref{eq:rapidFall}) for the case of a gradual end to
inflation with $a_{2f}/a_{1f} > 2$, we find (by blue shifting
the temperature at late times back in time)
that at the time
when the inflationary expansion is joined to the final
asymptotically flat segment of the expansion, the energy-density
has an effective temperature that is slightly larger than the
Gibbons-Hawking temperature \cite{Gibbons} of $H_{\rm infl}/2\pi$.
Particle creation by the expanding universe was also investigated
in \cite{Ford} as a possible cause of reheating, with a similar
result to the present one we find for a gradual end to inflation.

We next use our results to investigate the effective
temperature if there is a fairly abrupt end to inflation.
With a fairly abrupt end to inflation, for which
$a_{2f}/a_{1f} \simeq 1$, the range in $q_{2}$ where
$\left|\beta_{q_{2}}\right|^{2}\propto q_{2}{}^{-6}$ only begins
until a large value of $q_{2}$, which we denote
by $q_{2\rm cut-off}$. From our numerical results, we find that
$q_{2 \rm cut-off} \simeq a_{2f}/(a_{2f}-a_{1f})$.
We define the region between $1\lesssim
q_{2}\lesssim q_{2 \rm cut-off}$ as the ``stretched" region.  In this
``stretched" region,
$\left|\beta_{q_{2}}\right|^{2}\propto q_{2}{}^{-2}$.  As an example,
see Fig.~\ref{fig:massive}, where
$q_{2 \rm cut-off}\simeq10^{4}$ and
$a_{2f}/(a_{2f}-a_{1f})\simeq10^4$. The ``stretched" region is
a result of particle production caused by the rapid fall-off in
$H(t)$ as inflation ends.

With sufficient stretching, the particle number per
mode, $\left|\beta_{q_{2}}\right|^{2}$, in the
``stretched" region is proportional to $q_{2}^{\ -2}$ for any value
of $m_H < 3/2$.  In the intermediate-$q_{2}$ region, the
behavior of $\left|\beta_{q_{2}}\right|^{2}$ is governed by
the power $P$ defined in Eq.~(\ref{eq:P}) and shown in
Fig.~{\ref{fig:pvmh}}.
As is evident from Fig.~{\ref{fig:massive}}, the
stretched region extends over a much larger range of values
of $q_{2}$ than do the intermediate- and small-$q_{2}$ regions.
Because the $a(t)$ we defined is only $C^2$ across the joining
points, $\left|\beta_{q_{2}}\right|^{2}$ falls off as
$q_{2}{}^{-6}$ for $q_{2}> q_{2 \rm cut-off}$. For an $a(t)$ that
is $C^{\infty}$ for all $t$, the rate of fall off for
$q_{2}> q_{2 \rm cut-off}$ would be much faster. Therefore,
in our calculation below of the total energy density, we neglect
the contribution to the
energy density from this UV-range of $q_{2}$. (However, in the
$C^2$ case, one can show that the contribution of this UV-range to the total energy-density is of the same magnitude as the contribution
of the stretched region; so in that case the energy density would
be twice the value we obtain below.)

Therefore, for a fairly abrupt transition with
$a_{2f}/a_{1f} \simeq 1$, the total energy density is
\begin{eqnarray}
\left<\frac{E}{V}\right>&&\simeq\frac{1}{(2\pi
a_{2f})^{3}}\int_{a_{2f}H_{\rm infl}}^{a_{2f}H_{\rm infl}\ q_{2 \rm
cut-off}}\frac{k}{a_{2f}}\left|\beta_{k}\right|^{2}d^{3}k
\nonumber\\
&&=\int_{1}^{q_{2 \rm cut-off}}\frac{q_{2}^{\
3}\left|\beta_{q_{2}}\right|^{2}H_{\rm infl}^{4}}{2\pi^{2}}dq_{2}
\nonumber\\
&&=\int_{1}^{q_{2 \rm cut-off}}\frac{q_{2}H_{\rm
infl}^{4}}{8\pi^{2}}dq_{2}.
\end{eqnarray}
In the present case,
$q_{2 \rm cut-off}\simeq\left(a_{2f}/[a_{2f}-a_{1f}]\right)\gg 1$,
so we find
\begin{equation}
\left<\frac{E}{V}\right>\simeq\frac{H_{\rm
infl}^{4}\left(\frac{a_{2f}}{a_{2f}-a_{1f}}\right)^{2}}{16\pi^{2}}.
\end{equation}
It follows that a fairly abrupt end to inflation can lead to energy densities that correspond to an effective temperature large
with respect Gibbons-Hawking temperature.  This would
certainly be large enough to reheat the universe
if some process caused a fairly rapid end to inflation.
A more gradual end to inflation could also
produce a temperature sufficient for reheating if the
value of $H_{\rm infl}$ were large enough, as pointed out
in \cite{Ford}.

\section{Conclusions}

We have set up an exactly solvable model of exponential
inflation joined to initial and final asymptotically flat expansions
of the universe.  There are eight adjustable parameters, three of
which pertain to the initial asymptotically flat segment, three to the
final asymptotically flat segment, and two to the inflationary stage
of the expansion. In the case when the scale factor $a(t)$ is
required to be continuous with continuous first and second derivatives
at the joining times, there are three independent conditions on the
parameters, leaving five adjustable parameters.  Among these
are the rate of inflation, $H_{\rm infl}$, the number of e-foldings
of inflation, $N_e$, as well as three parameters characterizing
the asymptotic regions. The parameter $H_{\rm infl}$
is constant in this model, and $N_e$ can be arbitrarily large.
Our principal focus in this
work has been to use this exactly solvable model to look for
generic behaviors that have not been previously studied.
In this way, we have found a number of new effects that may
be relevant to the inflationary universe. We summarize these
as follows.

We find that if the quantized inflaton fluctuation field is
in its vacuum state in the early time asymptotic Minkowski space,
then the inflaton fluctations (which correspond to particles of a
quantized scalar field) are created by the expanding universe in
the manner
thoroughly analyzed in \cite{Parker2, Parker3}. We show that the creation of these particles, or inflaton fluctuations, leads to a
scale-invariant spectrum\footnote{We
are not incorporating the slow-roll
parameters, which of course would modify the scale-invariance.}
of created particles, or fluctuations,
for the range of momenta or wavelengths that are
relevant to the present universe when $N_e$ is
sufficiently large, for example, larger than about $60$.
For an initial Minkowski vacuum, we find that for modes having
wavelengths that are not so large that they leave the Hubble
sphere within the first few e-foldings of inflation, the Minkowski
vacuum state evolves into the Bunch-Davies de Sitter vacuum
after a few e-folds of inflation.  So the results for such modes
of the inflaton fluctuation field should agree with those
obtained by assuming the de Sitter vacuum, as is usually done
in treatments of inflation.

If the total number of e-foldings of inflation {\em prior} to the exit
of all the observationally relevant modes is not sufficiently long, then
our results would imply that the longest wavelength perturbations
that may come within the range of observation should have a
spectrum that deviates in a certain way from scale invariance.
This deviation is
a result of the fact that the Minkowski vacuum has not had time
to evolve to the de Sitter vacuum before the modes exit from
the Hubble sphere.

We find that if there were perturbations (i.e., particles)
present at early times, then
their effects can be propagated all the way
to the end of inflation, by means of ``stimulated emission" from
the vacuum.  These perturbations present prior to the start of
inflation would then have {\em observable} effects.
This is an important new effect that we are considering in a
paper\cite{GlenzParker} in preparation.

We investigated the spectrum of scalar perturbations of the
metric that would be induced by the inflaton fluctuations,
and we found the spectrum of initial density perturbation
that these would produce after the end of inflation.  We found
a significant difference between the properties of the
density perturbations that one would obtaiin by taking account
(a) only of the vacuum subtraction in the final Minkowski space,
as compared with (b) subtracting the full set of adiabatic regularization
terms at a time within a few e-foldings after a mode exits the
Hubble sphere during inflation, as in \cite{Agullo2}.
A new result we find is that in case (b), the density perturbations
have a non-zero value that is nearly independent of the inflaton
mass for masses $m_H \lesssim 0.25$.\footnote{At least for the quadratic inflaton potential that we considered, but probably also
more generally}. This type of behavior may permit one to use
our knowledge of the initial density perturbations to
make a prediction about the value of $H$ during inflation.
This mass-independence does not appear if one only uses
the MInkowski space vacuum subtraction.

We also considered the energy density that would be created
by the change of the scale factor $a(t)$ at the end of inflation.
We found agreement with earlier treatments for a gradual end
to inflation. We also were able to estimate the energy density
created if there were a fairly abrupt, but smooth, end to
inflation. The effective reheating temperature produced could
be very high. We give the value of the created energy density
as a function of a parameter characterizing the abruptness of
the end to inflation.

\begin{acknowledgments}
M.M.G. was supported by the Lynde and Harry Bradley Foundation,
and by the National Space Grant College and Fellowship Program and
the Wisconsin Space Grant Consortium.  L.P. has been partly
supported by NSF grants PHY-0071044 and PHY-0503366 and
by a UWM RGI grant.
\end{acknowledgments}

\begin{appendix}

\section{Boundary Matching Calculation}

\label{appendix:A}

Given the values of $\psi_{k1}$ and $\psi_{k1}'$, and the matching
conditions
\begin{eqnarray}
\label{eq:A1}
Ah_{1}(t_{1})+Bh_{2}(t_{1})&=&\psi_{k}(t_{1})=\psi_{k1}, \\
\nonumber Ah_{1}'(t_{1})+Bh_{2}'(t_{1})&=&\psi_{k}'(t_{1})=\psi_{k1}', \\
\nonumber Cg_{1}(t_{2})+Dg_{2}(t_{2})&=&Ah_{1}(t_{2})+Bh_{2}(t_{2}), \\
\nonumber
Cg_{1}'(t_{2})+Dg_{2}'(t_{2})&=&Ah_{1}'(t_{2})+Bh_{2}'(t_{2}),
\end{eqnarray}
we wish to calculate the constant coefficients $C$ and $D$ in terms
of the functions $h_{1}(t)$, $h_{2}(t)$, $g_{1}(t)$, and $g_{2}(t)$;
and the values of $\psi_{k1}$, $\psi_{k1}'$, $t_{1}$, and $t_{2}$.
(Here prime denotes derivative with respect to $t$.) Rearranging the
first two matching conditions leads to
\begin{eqnarray}
B=\left[\frac{\psi_{k1}-Ah_{1}}{h_{2}}\right]_{t=t_{1}}, \\
\nonumber
A=\left[\frac{\psi_{k1}'-Bh_{2}'}{h_{1}'}\right]_{t=t_{1}}.
\end{eqnarray}
Combining these two equations leads to
\begin{eqnarray}
A=\left[\frac{\psi_{k1}'h_{2}-\psi_{k1}
h_{2}'}{h_{1}'h_{2}-h_{1}h_{2}'}\right]_{t=t_{1}},
\nonumber\\
B=\left[\frac{\psi_{k1}'h_{1}-\psi_{k1}
h_{1}'}{h_{2}'h_{1}-h_{2}h_{1}'}\right]_{t=t_{1}}.
\end{eqnarray}
At the time, $t_{2}$, we have:
\begin{eqnarray}
\psi_{k}(t_{2})&=&Ah_{1}(t_{2})+Bh_{2}(t_{2})
\nonumber\\
&=&\bigg\{\left[\frac{\psi_{k1}'h_{2}-\psi_{k1}
h_{2}'}{h_{1}'h_{2}-h_{1}h_{2}'}\right]_{t=t_{1}}h_{1}(t_{2})
\nonumber\\
&&+\left[\frac{\psi_{k1}'h_{1}-\psi_{k1}
h_{1}'}{h_{2}'h_{1}-h_{2}h_{1}'}\right]_{t=t_{1}}h_{2}(t_{2})\bigg\},
\nonumber\\
\
\end{eqnarray}
and
\begin{eqnarray}
\psi_{k}'(t_{2})&=&Ah_{1}'(t_{2})+Bh_{2}'(t_{2})
\nonumber\\
&=&\bigg\{\left[\frac{\psi_{k1}'h_{2}-\psi_{k1}
h_{2}'}{h_{1}'h_{2}-h_{1}h_{2}'}\right]_{t=t_{1}}h_{1}'(t_{2})
\nonumber\\
&&+\left[\frac{\psi_{k1}'h_{1}-\psi_{k1}
h_{1}'}{h_{2}'h_{1}-h_{2}h_{1}'}\right]_{t=t_{1}}h_{2}'(t_{2})\bigg\}.
\nonumber\\
\
\end{eqnarray}
Let us also define $\psi_{k2}\equiv\psi_{k}(t_{2})$ and
$\psi_{k2}'\equiv\psi_{k}'(t_{2})$.  In terms of $\psi_{k2}$ and
$\psi_{k2}'$ the last two boundary conditions in Eq.~(\ref{eq:A1})
become
\begin{eqnarray}
C&=&\left(\frac{\psi_{k2}'g_{2}-\psi_{k2}
g_{2}'}{g_{1}'g_{2}-g_{1}g_{2}'}\right)_{t=t_{2}},
\nonumber\\
D&=&\left(\frac{\psi_{k2}'g_{1}-\psi_{k2}
g_{1}'}{g_{2}'g_{1}-g_{2}g_{1}'}\right)_{t=t_{2}}.
\end{eqnarray}
\\
Substituting for $\psi_{k2}$ and $\psi_{k2}'$ yields
\begin{eqnarray}
C=\left(\frac{[Ah_{1}'+Bh_{2}']g_{2}-[Ah_{1}+Bh_{2}]
g_{2}'}{g_{1}'g_{2}-g_{1}g_{2}'}\right)_{t=t_{2}},
\nonumber\\
D=\left(\frac{[Ah_{1}'+Bh_{2}']g_{1}-[Ah_{1}+Bh_{2}]
g_{1}'}{g_{2}'g_{1}-g_{2}g_{1}'}\right)_{t=t_{2}}.
\end{eqnarray}
Finally, expressing $A$ and $B$ in terms of the given values of
$\psi_{k1}$ and $\psi_{k1}'$ specified at $t_{1}$ leads to
\begin{eqnarray}
C&=&\frac{1}{\left(g_{1}'g_{2}-g_{1}g_{2}'\right)_{t=t_{2}}}
\nonumber\\
&&\times\bigg\{\left[\frac{\psi_{k1}'h_{2}-\psi_{k1}
h_{2}'}{h_{1}'h_{2}-h_{1}h_{2}'}\right]_{t=t_{1}}\left(h_{1}'g_{2}-h_{1}g_{2}'\right)_{t=t_{2}}
\nonumber\\
&&+\left[\frac{\psi_{k1}'h_{1}-\psi_{k1}
h_{1}'}{h_{2}'h_{1}-h_{2}h_{1}'}\right]_{t=t_{1}}\left(h_{2}'g_{2}-h_{2}g_{2}'\right)_{t=t_{2}}\bigg\},
\nonumber\\
\
\end{eqnarray}
and
\begin{eqnarray}
D&=&\frac{1}{\left(g_{2}'g_{1}-g_{2}g_{1}'\right)_{t=t_{2}}}
\nonumber\\
&&\times\bigg\{\left[\frac{\psi_{k1}'h_{2}-\psi_{k1}
h_{2}'}{h_{1}'h_{2}-h_{1}h_{2}'}\right]_{t=t_{1}}\left(h_{1}'g_{1}-h_{1}g_{1}'\right)_{t=t_{2}}
\nonumber\\
&&+\left[\frac{\psi_{k1}'h_{1}-\psi_{k1}
h_{1}'}{h_{2}'h_{1}-h_{2}h_{1}'}\right]_{t=t_{1}}\left(h_{2}'g_{1}-h_{2}g_{1}'\right)_{t=t_{2}}\bigg\},
\nonumber\\
\
\end{eqnarray}
which are the combined joining conditions for $\psi_{k}$ and
$\psi_{k}'$.

\end{appendix}


\begin{thebibliography}{9}

\bibitem{SR}
A. H. Guth,  {\it Phys. Rev. D} {\bf 23}, 347, (1981); K. Sato, {\it
Phys. Lett.} {\bf 99B}, 66, (1981); A. A. Starobinsky, {\it Phys.
Lett.} {\bf 91B}, 99 (1980); A. A. Starobinsky, {\it Phys. Lett.}
{\bf 117B}, 175 (1982); A. Guth and S.-Y. Pi, Phys. Rev. Lett.
\textbf{49}, 1110 (1982); J.M. Bardeen, P.J. Steinhardt, and
M.S.Turner, Phys. Rev. \textbf{D 28}, 679 (1983); R. Brout, F.
Englert, and E. Gunzig, {\it Ann. of Phys.} {\bf 115}, 78 (1978);
V.F. Mukhanov, H.A. Feldman, and R.H. Brandenberger, Phys. Rept.
\textbf{215}, 203 (1992).

\bibitem{Parker1} L. Parker, hep-th/0702216 (2007).

\bibitem{BiLuPiJuAn}
N. D. Birrell, {\it Proc. Roy. Soc. (London)\/},
{\bf A361}, 315 (1978);
C. L\"{u}ders, and J. E. Roberts,
{\it Commun. Math. Phys.}{\bf 134}, 29 (1990);
K. Pirk, {\it Phys.  Rev. D} {\bf 48}, 3779 (1993);
W. Junker, and E. Schrohe,
Annales Poincare Phys. Theor. {\bf 3}, 1113 (2002);
arXiv:math-ph/0109010.

\bibitem{Agullo} I. Agull\'{o}, J. Navarro-Salas, G.J. Olmo, and L.
Parker, Phys. Rev. Lett. \textbf{101}, 171301 (2008).

\bibitem{Agullo2} I. Agull\'{o}, J. Navarro-Salas, G.J. Olmo, and L.
Parker, ``Revising the predictions of inflation for the cosmic
background anisotropies," arXiv:0901.0439 (2009).

\bibitem{Parker2}
L. Parker, \textit{The creation of particles by the
expanding universe}, Ph.D. thesis, Harvard University (1966).

\bibitem{Parker3}
L. Parker, {\it Phys. Rev. Lett.} {\bf 21}, 562 (1968); See also L.
Parker,  {\it Phys. Rev.} {\bf 183}, 1057, (1969), where further
results of \cite{Parker2} were presented, including the connection
between conformal invariance and particle creation. The
``gravitons'' referred to there were quanta satisfying the
conformally invariant spin-2 quantized field equation, as one can
see from their given field equations. The reader should note that
the gravitons obtained by linearizing the Einstein gravitational
field equation are not conformally invariant. These Einstein
gravitons are created in exactly the same way as the minimally
coupled scalar particles that were shown to be created in
\cite{Parker2, Parker3}.  [The linearized Einstein graviton field
equations in the Lifshitz gauge of an FRW universe, where the two
independent polarizations each satisfy a field equation identical to
the minimally-coupled scalar field equation, are given  in E. M.
Lifshitz, {\it Zh. Eksp. Teor. Fiz.}{\bf 16}, 587 (1946); and are
discussed with respect to particle creation in L. P. Grishchuk, {\it
Zh. Eksp. Teor. Fiz.}{\bf 67}, 825 (1974); {\it Sov.
Phys.--JETP}{\bf 40}, 409 (1975); and L. H. Ford and L. Parker, {\it
Phys. Rev. D}{\bf 16}, (1977). ]

\bibitem{Parker5}
L. Parker, L. and S. A., Fulling, {\it Phys. Rev. D} {\bf 9}, 341
(1974); S. A., Fulling, L. Parker, and B. L. Hu, {\it Phys. Rev. D}
{\bf 10}, 3905 (1974); see also P. R. Anderson, and L. Parker, {\it
Phys. Rev. D} {\bf 36}, 2963 (1987).

\bibitem{Allen} B. Allen, Phys. Rev. \textbf{D 37}, 2078 (1988).

\bibitem{Yajnik} U.A. Yajnik, Phys. Lett. \textbf{B 234}, 271
(1990).

\bibitem{3WMAP} D.N. Spergel, et.al., ApJS \textbf{170}, 377 (2007).

\bibitem{5WMAP} E. Komatsu, et.al., ApJS \textbf{180}, 330 (2009).

\bibitem{Ford} L.H. Ford, Phys. Rev. \textbf{D 35}, 2955 (1987).

\bibitem{EE} P.J. Epstein,  \emph{Proc. Nat. Acad. Sciences (US)}
\textbf{16}, 627 (1930); C. Eckart, Phys. Rev. \textbf{35}, 1303
(1930).

\bibitem{CinNature} L. Parker, Nature \textbf{261}, 20
(1976); L. Parker, ``The Production of Elementary Particles in
Strong Gravitational Fields,'' in \emph{Asymptotic Structure of
Space-Time}, edited by F.P. Esposito and L. Witten, (Plenum Press,
New York), 107 (1977); L. Parker, ``Quantized Fields and Particle
Creation in Curved Spacetime,'' 66 pages in {\it Relativity, Fields,
Strings and Gravity: The Second Latin American Symposium on
Relativity and Gravitation (SILARG 2)}, editor C. Aragone.
(Universidad Simon Bolivar, Caracas, 1975).

\bibitem{ParkerToms} L. Parker and D.J. Toms,
\textit{Quantum Field Theory in Curved Spacetime:
quantized fields and gravity} (Cambridge University Press, 2009).

\bibitem{AbramowitzStegun} M. Abramowitz and I.A. Stegun,
\emph{Handbook of Mathematical Functions}
(U.S. Dept. of Commerce : U.S. G.P.O., 1972).

\bibitem{AllenFolacci}
B. Allen and A. Folacci, Phys. Rev. \textbf{D 35}, 3771 (1987).

\bibitem{Kulsrud}
R.M. Kulsrud, Phys. Rev. \textbf{106}, 205 (1957);
J.E.Littlewood, Ann. Phys. (NY) {\bf 21}, 233 (1963).

\bibitem{Chung}
D.J.H. Chung,
E.W. Kolb, and A. Riotto, Phys. Rev. \textbf{D 59}, 023501
(1999), see their Appendix; also see \cite{Parker2}.

\bibitem{ParkerNY}
L. Parker, ``Time's Arrow and the Strength of Inflation,'' talk
presented at the Origins of Time's Arrow conference at the New York
Academy of Sciences, October 15-16 (2007).

\bibitem{Anderson} P.R. Anderson, C. Molina-Par\'{\i}s, and E.
Mottola, Phys. Rev. \textbf{D 72}, 043515 (2005).

\bibitem{LL}
A.R. Liddle and D.H. Lyth, \textit{Cosmological inflation and
large-scale structure} (Cambridge University Press, 2000).

\bibitem{bunch-davies}
T.S. Bunch and P.C.W. Davies, Proc. Roy. Soc. London \textbf{A 360},
117 (1978).

\bibitem{Andersonetal}
P.R. Anderson, W. Eaker, S. Habib, C. Molina-Paris, and E. Mottola,
Phys. Rev. \textbf{D 62}, 124019 (2000).

\bibitem{Dodelson}
S. Dodelson, \textit{Modern Cosmology} (Academic Press, 2003).

\bibitem{Kiefer} Kiefer C., Lohmar I., Polarski D.,
Starobinsky A. A.,
Class. Quant. Grav. 24 1699 (2007); astro-ph/0610700v2.

\bibitem{KT} E.W. Kolb and M.S. Turner, \textit{The Early Universe}
(Perseus Publishing, 1994).

\bibitem{Linde}
A.D. Linde, Phys. Lett. 108B, 389 (1982);
S. Habib, A. Heinen, K. Heitmann, and G. Jungman,
Phys. Rev. \textbf{D 71}, 043518 (2005).

\bibitem{Dunkley} J. Dunkley, et.al., ApJS \textbf{180}, 306 (2009).

\bibitem{Gibbons} G.W. Gibbons and S.W. Hawking, Phys. Rev.
\textbf{D 15}, 2738 (1977).

\bibitem{GlenzParker} M.M. Glenz and L. Parker, ``Propagation of
early perturbations through arbitrarily long inflation
without depletion," ``Observable effects that survive inflation,''
in preparation (2009).


\end{thebibliography}
\end{document}